\documentclass[11pt,letterpaper]{JHEP3}
\usepackage{amssymb}

\newcommand{\be}{\begin{equation}}
\newcommand{\ee}{\end{equation}}
\def\bea{\begin{eqnarray}}
\def\eea{\end{eqnarray}}
\def\der{\partial}
\newcommand{\hh}{\ensuremath{{\hat{h}}}}
\newcommand{\M}{\ensuremath{\mathcal{M}}}

\newcommand{\mscr}[1]{\mbox{\scriptsize #1}}

\newcommand{\ft}[2]{{\textstyle\frac{#1}{#2}}}

\title{Gauged Supergravity and Singular Calabi-Yau Manifolds}
\author{Thomas Mohaupt\thanks{New address after October 21, 2001: 
Theoretisch-Physikalisches Institut, Friedrich-Schiller-Universit\"{a}t
Jena, Max-Wien-Platz 1, D-07743 Jena, Germany.} { } and
Marco Zagermann\thanks{Work supported  by
the ``Schwerpunktprogramm 1096'' of 
the German Science Foundation (DFG)}\\Fachbereich Physik,
Martin-Luther-Universit\"{a}t Halle-Wittenberg,\\
Friedemann-Bach-Platz 6, D-06099 Halle, Germany\\ E-mail: 
\email{mohaupt@physik.uni-halle.de, zagermann@physik.uni-halle.de}}
\abstract{Compactifications of M-theory on singular manifolds 
contain additional
charged massless states descending from M-branes wrapped on
vanishing cycles. We construct the   first  explicit  example of a
complete supergravity
Lagrangian that includes such extra states.
This is done for a  compactification on a Calabi-Yau threefold
that  develops a genus
zero curve of $A_1$ singularities at the boundary of the K\"ahler cone
with a resulting   $SU(2)$ gauge symmetry enhancement.
The corresponding $SU(2)$ gauged supergravity Lagrangian
includes  two charged  and two neutral vector multiplets, and
turns out to be  uniquely fixed
by the Calabi-Yau geometry and by the effective ungauged Lagrangian
describing
the Coulomb branch. One can see
explicitly how resolving the singularity corresponds to a supersymmetric
Higgs effect in the gauged supergravity description.
The elementary transformation relating the
two families of smooth Calabi-Yau resolutions of the singularity
acts as the $SU(2)$ Weyl twist.
The resulting structure appears to be very rigid and is likely
to apply to other types of singularities and manifolds as well.}
\preprint{hep-th/0109055}

\begin{document}

\renewcommand{\theequation}{\arabic{section}.\arabic{equation}}
\section{Introduction}
\setcounter{equation}{0}

One of the most remarkable properties of string theory is that it
can make sense of certain singular space-time backgrounds. The
most prominent examples in the context of Calabi-Yau
compactifications are flop transitions \cite{EW93},
conifold singularities \cite{Str95,GMS} and
their generalizations \cite{KM96,KMP}. If the singularity of the
Calabi-Yau space can be resolved in more than one way,
transitions between spaces of different topology are possible.
Many examples of this have been discussed not only for
type II string theory, but also for M-theory and F-theory
compactifications \cite{MV96,EW96,MS}. The basic mechanism that leads to
regular physics in an apparently singular space-time involves the
presence of branes which  can wrap around cycles. In the limit
where such a cycle becomes singular, the brane contributes
additional light degrees of freedom to the low energy spectrum.
An effective supergravity action which only contains those
modes that are massless for generic moduli becomes singular in
this region of moduli space, whereas the effective action including the
additional light modes is expected to be regular.

The intuitive picture behind this is that, in order to really
observe the singular behaviour of such a  background geometry, a physical
observer
would have to probe the singularity with test particles (or test
strings or test branes). Such observable singularities would for
example be a feature of ordinary supergravity compactifications on
singular Calabi-Yau spaces. But once brane degrees of freedom in
the full string or M-theory are taken into account, any test
particle used to probe the would-be singularity instead interacts
with the brane wrapped around the vanishing cycle, with the result
that all physical observables are regular.

Apart from their importance for a deeper understanding of geometry
in a theory of quantum gravity, singular Calabi-Yau spaces and
topological transitions have also been used to study
non-perturbative aspects of supersymmetric field theories \cite{KKLMV,MS}.
Moreover, branes wrapping vanishing cycles have appeared in
recent proposals for solving the gauge hierarchy problem \cite{GKP}
(using a compact version of the Klebanov-Strassler solution \cite{KS}),
the problem
of the cosmological constant \cite{FMRSW} and the cosmological moduli problem
\cite{Dine}. It is therefore desirable to have a complete
understanding of the theory including the extra light modes.

If one takes  a closer look at the existing literature, however,  one
finds that one important ingredient still seems to be missing,
namely the explicit specification of a
full effective Lagrangian that also encompasses the dynamics of
the extra light modes. Instead,
the reasoning usually goes along the lines of Strominger's
classical paper on the conifold
singularity \cite{Str95}, that is, it is\\
(i) shown that a singularity appearing in the effective action
without the extra light modes can be interpreted as arising from
illegitimately integrating out a massless supermultiplet of
specific type and charge, and\\
(ii) it is shown that the corresponding model contains branes
wrapped on vanishing cycles which give rise to precisely this type
of multiplet.

Using (algebraic) geometry both qualitative and quantitative
understanding of the dynamics at these special loci in moduli
space has been gained. But to our knowledge a full and manifestly
regular effective supergravity Lagrangian capturing all relevant
low energy physics at the singular point has not been written down
in any of the cases. From a conceptual point of  view,   this is
of course not very satisfactory. Moreover, in many applications
that involve singular background manifolds,
one always has to work with a singular Lagrangian. Although one
believes to understand the nature of the singularity and its cure,
such results invite criticism, and it is preferable to have a
regular Lagrangian including all relevant modes as a solid
starting point.

One should also be aware that a fundamental microscopic
description of such  singular compactifications is not available
in many cases. Indeed, for M-theory compactifications one only
knows about the eleven-dimensional theory that it has supergravity
as its low energy limit and contains BPS solitons such as the
M2-brane and the M5-brane, whereas for type II compactifications
the conformal field theory description becomes singular for the
conifold and its generalizations. This makes a macroscopic
description by a supergravity Lagrangian even more valuable.

In this paper we take a first step towards constructing such
Lagrangians. We will consider a specific case which can be brought
under full control, namely the vector multiplet sector of
five-dimensional, $\mathcal{N}=2$  \footnote{By
``$\mathcal{N}=2$'' we mean the minimal amount of supersymmetry
in five space-time dimensions corresponding to eight real
supercharges.} supergravity \cite{GST1,GST2,GZ1}.
Such theories can be obtained by
compactification of eleven-dimensional supergravity \cite{CJS} on a
Calabi-Yau threefold \cite{CY,AFT}, and when this threefold develops a specific
type of singularity --- a genus zero curve of $A_1$ singularities
--- one observes non-Abelian gauge symmetry enhancement $U(1)
\rightarrow  SU(2)$ without additional matter (i.e.,
hypermultiplets) \cite{EW96}. Since the vector multiplet sector is uniquely
specified in terms of a cubic polynomial (the `prepotential')
\cite{GST1}, the problem of finding the $SU(2)$ gauged
supergravity Lagrangian that includes the two additional charged
vector multiplets is tractable, and we will show that this
extended Lagrangian is
uniquely determined by the generic low energy effective action and
the Calabi-Yau geometry.

Although other and more interesting situations such as
five-dimensional supergravity coupled to both vector and
hypermultiplets,  Calabi-Yau compactifications with background
flux or type II compactifications to four dimensions can be
considerably more complicated, we obtain already plenty of
interesting and encouraging results in our simplest example. Most
importantly, we find that a successful reconstruction of the full
$SU(2)$ gauged Lagrangian requires a careful consideration of the
one-loop threshold effects described in \cite{AFT,EW96,MS}.
Without these threshold corrections no consistent embedding into
an $SU(2)$ gauged supergravity Lagrangian with the right matter
content is possible. On the other hand, if one does take these
corrections into account (with precisely the correct coefficient),
the corresponding $SU(2)$ gauged Lagrangian turns out to be
uniquely determined, i.e.,  no free parameter is introduced by the
coupling of the two additional vector multiplets, as one might
perhaps have expected naively.

As another interesting result, we will get a gauge theory picture of so-called
elementary transformations of Calabi-Yau spaces in terms of the
$SU(2)$ Weyl twist. At least a posteriori we realize that the
problem of reconstructing the coupling to the two
charged vector multiplets is dictated by $SU(2)$ gauge symmetry. A
similar structure seems to be present for flop transitions, though
in that case it is not related to gauge symmetry enhancement. This
makes us hope that systematic rules govern the re-implementation
of extra light modes.

The organization of this paper is as follows. In Section 2, we
briefly summarize the relevant structures of ungauged 5D,
$\mathcal{N}=2$ supergravity theories and  recall how they descend
from (smooth) Calabi-Yau compactifications of 11D supergravity. In
Section 2.3, this general discussion is then illustrated with the
specific examples we consider in this paper. The Calabi-Yau
picture of these models suggests that at certain points in the moduli space,
at which the Calabi-Yau space becomes singular, an
$SU(2)$ gauge symmetry enhancement takes place. In Section 3, the
main part of this paper, we then attempt a construction of the
underlying $SU(2)$ gauged supergravity Lagrangian that also
includes the additional light modes responsible for the gauge
symmetry enhancement. In Section 4, we extract a relation between
the elementary transformations of the Calabi-Yau spaces and the
$SU(2)$ Weyl twist from our model. A summary of our results and a
list of possible applications and some interesting generalizations
are given in Section 5. Appendices A and B contain some  details
on the calculations described in Section 3.

Effective Lagrangians which include
additional modes that only
become massless at special points in the moduli space have been
discussed before in the literature (see \cite{GPR} for a review).
Our results extend
these in various directions:\\
(i) We are not restricted to stringy perturbative mechanisms of
gauge symmetry enhancement (though these are also covered) but can
also include additional states coming from branes wrapped on
vanishing cycles.\\
(ii) All the existing examples work either in situations where the
moduli space is a symmetric space and no threshold effects are
present, i.e., in $\mathcal{N}\geq 4$ compactifications \cite{GPR},
or\\
(iii) these models are explicitly considered
 only in such a (classical) approximation. An example for the latter are
    the untwisted
sectors of orbifolds studied in \cite{ILLT}.

\section{Ungauged five-dimensional supergravity and smooth Calabi-Yau
threefolds}
\setcounter{equation}{0}
The coupling of an arbitrary number of vector and hypermultiplets
to 5D, $\mathcal{N}=2$ supergravity in the absence of any gauge
interactions was constructed in \cite{GST1,Sier}. The
relation to Calabi-Yau compactifications of 11D supergravity
was explicitly worked out in
\cite{CY,AFT}. We refer to these references for further details.

\subsection{Maxwell-Einstein supergravity}
The effective field theory of the compactification of
eleven-dimensional supergravity on a smooth Calabi-Yau threefold,
$X$, is described by minimal five-dimensional supergravity coupled
to $n_V$ Abelian vector multiplets and $n_H$ neutral
hypermultiplets. The hypermultiplets will play no r\^{o}le
in the following, though we will make some remarks on
generalizations involving hypermultiplets in Sections 4 and 5. \\
The supermultiplets we are concerned with are thus\\

\noindent
(i) The supergravity multiplet:
\be (e_{\mu}^{\;\;m},
\psi^i_{\mu}, A_{\mu})
\ee
consisting of the graviton
$e_{\mu}^{\;\;m}$, two gravitini $\psi^{i}_{\mu}$ and one vector
field $A_{\mu}$.\\

\noindent
(ii) The vector multiplet: \be (A_{\mu}, \lambda^{i}, \phi) \ee
comprising one vector field $A_{\mu}$, two spin-1/2 fermions (gaugini)
$\lambda^{i}$ and one real scalar field $\phi$.

Here, $\mu,\nu = 0,\ldots, 4$ and $m,n =0,\ldots,4$ are,
respectively,  curved and flat space-time indices, whereas $i=1,2$
is a doublet index associated with the automorphism group
$USp(2)_R \simeq SU(2)_R$ of the underlying supersymmetry algebra.
All fermions
obey a symplectic Majorana condition with respect to that index
$i$ (see \cite{GST1} for more details).

Coupling  $n_V$ Abelian vector multiplets to supergravity then
results in the total field content  \be (e_{\mu}^{\;\;m},
\psi^i_{\mu}, A_{\mu}^I, \lambda^{ix}, \phi^x), \ee where we have
combined the graviphoton with the $n_V$ vector fields from the
vector multiplets to form a single $(n_V+1)$-plet of vector fields
$A_{\mu}^{I}$ labelled by the index $I=0,1,\ldots,n_V$. The index
$x=1,\ldots,n_V$ is a curved index on the  $n_V$-dimensional
Riemannian target space, $\mathcal{M}$, of the scalar fields
$\phi^x$. The metric on $\M$ will be denoted by $g_{xy}$.

Introducing the Abelian field strengths $F_{\mu\nu}^{I}\equiv 2
\partial_{[\mu}A_{\nu]}^{I}$, the bosonic part of the Lagrangian
reads \cite{GST1}

\begin{eqnarray}\label{Lagrange1}
e^{-1}\mathcal{L}_{bosonic}&=&-\frac{1}{2}R- \frac{1}{4}
{\stackrel{\circ}{a}}_{IJ} F_{\mu\nu}^{I}F^{\mu\nu J}- \frac{1}{2}
g_{xy}(\partial_{\mu}\phi^{x})(\partial^{\mu}
\phi^{y})\nonumber\\
&& +\frac{e^{-1}}{6\sqrt{6}}C_{IJK}
\varepsilon^{\mu\nu\rho\sigma\lambda}F_{\mu\nu}^{I}
F_{\rho\sigma}^{J}A_{\lambda}^{K}.
\end{eqnarray}

The completely symmetric tensor $C_{IJK}$ in the $FFA$ term of
(\ref{Lagrange1}) is independent of the scalar fields and
completely determines the entire theory \cite{GST1}. To be more
explicit, the $C_{IJK}$ define a cubic polynomial
\begin{equation}
\mathcal{V}(h):=C_{IJK}h^{I}h^{J}h^{K}
\end{equation}
in $(n_V+1)$ real variables $h^{I}$ ($I=0,\ldots, n_V$), which
endows ${\mathbb{R}}^{(n_V+1)}$ with the  metric
\begin{equation}\label{aij}
a_{IJ}(h):=-\frac{1}{3}\frac{\partial}{\partial h^{I}}
\frac{\partial}{\partial h^{J}} \ln \mathcal{V}(h).
\end{equation}
The $n_V$-dimensional `very special'  manifold $\M$
(see \cite{dWvP2} for the origin
of this name) can then be
represented as the hypersurface \cite{GST1}
\begin{equation}
{\cal V} (h)=C_{IJK}h^{I}h^{J}h^{K}=1.
\end{equation}
This hypersurface constraint can (locally) be solved for the
physical scalars $\phi^x$. One convenient way is to take the
independent ratios (special coordinates), i.e., \be \phi^x =
\frac{h^x}{h^0} \ee in the coordinate patch $h^0 \not=0$, but any
other coordinate choice is of course also possible.

The  metric $g_{xy}$ on $\M$ is then simply the induced metric of
(\ref{aij}),
\begin{displaymath}
g_{xy}(\phi) = \frac{3}{2}\left. \frac{\der h^I}{\der \phi^x}
\frac{\der h^J}{\der \phi^y} a_{IJ} \right|_{ {\cal V} = 1} \;,
\end{displaymath}
whereas ${\stackrel{\circ}{a}}_{IJ}(\phi)$ is given by the
restriction of $a_{IJ}$ to $\M$,
\be
{\stackrel{\circ}{a}}_{IJ}(\phi)=a_{IJ}|_{{\cal V}=1}\;.
\label{a0}
\ee

For later reference, we define
\be \label{hIx}
h^I_x := -
\sqrt{\frac{3}{2}} \frac{\der h^I}{\der \phi^x}  \;.
\ee

\subsection{Calabi-Yau compactification}

When the above supergravity theories are obtained from a
compactification of 11D supergravity on a smooth Calabi-Yau
threefold $X$, then the numbers of vector and hypermultiplets are
given by \be n_V = h^{1,1}(X) -1 \;, \;\;\; n_H = h^{2,1}(X) + 1,
\ee where $h^{p,q}(X)$ are the Hodge numbers of $X$. The
Calabi-Yau space has $h^{1,1}(X)$ real K\"ahler moduli and
$h^{2,1}(X)$ complex moduli, which parameterize deformations of the
complex structure. The vector multiplet moduli are all the
K\"ahler moduli except the one parameterizing the overall volume.
The hypermultiplet moduli space contains the overall volume of the
Calabi-Yau space, the complex  structure moduli, and the
moduli obtained by dimensional reduction of the three-form gauge
field.

In a Calabi-Yau compactification the embedding coordinates $h^{I}$
can be chosen such that they parameterize the volumes of the
homological two-cycles of $X$. If $C^I$ is a basis of
$H_2(X,{\mathbb{Z}})$, and $J$ is the K\"ahler form of $X$, then
\be h^I = \int_{C^I} J = \mbox{vol}(C^I) \ee (where $\mbox{vol}$
is understood to be the volume of a supersymmetric (i.e.
holomorphic) curve in the homology class $C^I$. Such a curve has
minimal volume in $C^{I}$.) In this basis it is obvious that one
needs to have $h^I>0$ so that the volumina of all two-cycles are
postive. This defines the so-called K\"ahler cone. At the boundary
of the K\"ahler cone, defined by $h^{I} = 0$ for some $I$, the
corresponding cycle collapses and the manifold $X$ becomes
singular.

The coefficients $C_{IJK}$ have the interpretation as the triple
intersection numbers of $X$. Let $D_I$ be a basis of
$H_4(X,{\mathbb{Z}})$ (the four-cycles) that is dual to the
basis $C^I$ in the sense that
\be C^I \cdot D_J = \delta^I_J \;,
\ee
where $\cdot$ denotes the intersection product. Then
\be C_{IJK} =
D_I \cdot D_J \cdot D_K = \int_X
\omega_I \wedge \omega_J \wedge \omega_K \;,
\ee
where the two-forms $\omega_I$ are the Poincar\'{e} duals of
the four-cycles $D_I$. They form a basis of $H^2(X,{\mathbb{Z}})$.
The Poincar\'{e} duals $\omega^I$ of the two-cycles $C^I$ form a basis
of  $H^4(X,{\mathbb{Z}})$.

The volumes of four-cycles $D_I$  are given by
\be
\mbox{vol}(D_I) = \int_{D_I} J \wedge J = h_I \;,
\ee
where we expanded the K\"ahler form $J$ in the basis $\omega_I$
and introduced the dual embedding
coordinates $h_I$ defined by
\be h_I = C_{IJK} h^J h^K.
\label{hdual}
\ee

The volume of the Calabi-Yau space $X$ is \be \mbox{vol}(X) =
\int_X J \wedge J \wedge J = C_{IJK} h^I h^J h^K \;. \ee Since the
modulus corresponding to the total volume $\mbox{vol}(X)$ does not
sit in a vector multiplet but in a hypermultiplet, one needs to
introduce rescaled fields $h'^I = ( \mbox{vol}(X) )^{-1/3} h^I$ in
order to disentangle vector multiplet moduli and hypermultiplet
moduli \cite{CY}. These parameterize a hypersurface in the
K\"ahler cone, which is just the vector multiplet moduli space:
\be {\cal V}(h') = C_{IJK} h'^I h'^J h'^K = 1 \;. \ee For
notational convenience and to be consistent with the notation used
in Section 2.1 we will drop the primes from now on, $h'^I
\rightarrow h^I$. It is also understood that volumes of cycles and
the K\"ahler form $J$ have been rescaled with appropriate powers
of the total volume. Thus when talking about the volume of a cycle
we actually mean the volume measured in units of the overall
volume of the Calabi-Yau space.

The BPS-extended $D=5,{\cal N}=2$
supersymmetry algebra contains a (scalar) electric central
charge $Z_{(e)}$ and a (vectorial) magnetic central charge
$Z_{(m)}$. To these correspond BPS bounds for the masses of
electrically charged point particles and the tensions of
magnetically charged strings. These central charges can be
expressed as \cite{AFT} \be Z_{(e)} = q_I h^I \mbox{   and }Z_{(m)} =
p^I h_I \ee where $q_I, p^I$ are the electric and magnetic charge
with respect to the vector fields
$A^I_{\mu}$. In Calabi-Yau compactifications one
can easily identify BPS solitons which saturate these bounds \cite{EW96}.
Wrapping an M2-brane on the two-cycle $C^I$ gives pointlike sates
with charge $q_I= \pm 1 $. In order to get a BPS state, one needs
to wrap around a holomorphic curve in this class, which then has
volume $h^I$. The mass of such a state is \be
\label{BPSmass}
 M = T_{(2)}
\mbox{vol}(C^I) = T_{(2)} h^I \;.
\ee
It saturates the BPS bound
$M = T_{(2)} |Z_{(e)}|$, where $T_{(2)}$ denotes
the tension of the M2-brane.
There are also states which carry charge with respect to several
$U(1)$s. These descend from branes which wrap non-trivially around
several generators of $H_2(X,\mathbb{Z})$. In an analogous way,
the M5 brane can be wrapped on four-cycles resulting in a
magnetic string with charges $p^I$, where $p^I$ are the expansion
coefficients of the cycle in the basis of $H_4(X,\mathbb{Z})$.
The tension of such a string is $T = T_{(5)} |Z_{(m)}|$, where
$T_{(5)}$ is the tension of the M5-brane.

Note that at the boundary of the K\"ahler cone (i.e., when
$h^{I}\rightarrow 0$) one always gets additional charged massless
states descending from wrapped M2-branes. The precise nature of
these states depends on the details of the collapsing cycle. Two cases
which generically appear at the boundary of the K\"ahler cone
are well-understood \cite{EW96}.
\begin{enumerate}
\item
The homology class $C^I$ contains a finite number, $N$, of
isolates holomorphic curves. In this case,  one obtains $N$
charged hypermultiplets.
\item
The homology class $C^I$ contains a continuous family of
holomorphic curves with the following properties: the family is itself
parameterized by a holomorphic curve of genus $g$ and therefore
sweeps out a divisor $E$, which defines a class in
$H_4(X,{\mathbb{Z}})$. At the boundary $h^{I}=0$ the divisor $E$
collapses into a curve of $A_1$ singularities. (This means that,
when the singular curve is intersected with a complex surface, the
local geometry around the singular point is
${\mathbb{C}}^2/{\mathbb{Z}}_2$.) In terms of homology, the
collapse of the two-cycle $C^I$ induces the collapse of the
four-cycle $E$ into a two-cycle. In this case, one obtains two
charged vector multiplets, which combine with the vector multiplet
containing $A^I_{\mu}$ to form the adjoint representation of
$SU(2)$. Thus the (inverse of the) Higgs effect occurs at the
boundary of the K\"ahler cone.
In addition to the two charged vector multiplets,
one also gets $2g$ charged
hypermultiplets which combine with $g$ uncharged hypermultiplets \cite{KMP} to
form $g$ adjoint hypermultiplets.
\end{enumerate}
The additional charged massless modes can be identified
by a collective mode analysis
in eleven-dimensional supergravity \cite{EW96}.

Our aim is to construct the gauged supergravity Lagrangians that
include the additional light charged multiplets. In this paper, we
consider the case where only vector multiplets become massless,
and where these additional vector multiplets lead to an $SU(2)$
gauge symmetry enhancement. The corresponding singularity in $X$
is thus   a genus zero curve of $A_1$ singularities.

\subsection{Our models}

We  will consider three Calabi-Yau compactifications in parallel,
which all give rise to two vector multiplets and exhibit $SU(2)$
gauge symmetry enhancement at one boundary of the K\"ahler cone
without additional hypermultiplets.
The corresponding Calabi-Yau
manifolds $X_n$, $(n=0,1,2)$ are elliptic fibrations over the
Hirzebruch surfaces ${\bf F}_n$. An explicit construction as
hypersurfaces in toric varieties can be found in \cite{LSTY}.
The prepotentials are \bea {\cal V}_{(0)} &=&
8 (h^0)^3 + 6 (h^0)^2 h^1 + 6 (h^0)^2 h^2 + 6 h^0 h^1 h^2 \\
{\cal V}_{(1)} &=&
8 (h^0)^3 + 9 (h^0)^2 h^1 + 3 h^0 (h^1)^2 + 6 (h^0)^2 h^2 + 6 h^0 h^1 h^2 \\
{\cal V}_{(2)} &=& 8 (h^0)^3 +  12 (h^0)^2 h^1 + 6 (h^0)^2 h^2 + 6
h^0 h^1 h^2  \eea with the corresponding K\"ahler cones $h^I > 0$.

It is convenient to introduce new embedding coordinates $S,T,U$
defined by \bea
6^{1/3} h^0  &=& U \\
6^{1/3} h^1  &=& T-U \\
6^{1/3} h^2  &=& \left\{ \begin{array}{ll}
S - U & \mbox{   for   } n=0 \\
S - \ft12 (T+U) & \mbox{   for  }n=1 \\
S- T & \mbox{   for  } n=2. \\
\end{array} \right. \\
\eea Then all three prepotentials take the same form \be {\cal V}
= STU + \ft13 U^3,
 \ee while the K\"ahler cones are now defined by
\be T > U  > 0 \ee and \bea
S > U & & \mbox{for  } n=0\\
S > \ft12 (T+U) & & \mbox{for  } n=1 \\
S > T & & \mbox{for  } n=2.  \eea This choice of variables is
natural in the dual heterotic description of the model.
Compactification of M-theory on $X_n$ is dual to
compactifications of the heterotic string on $K3 \times S^1$ with
instanton numbers $(12+n,12-n)$ \cite{AFT,MV96,EW96,LSTY}.

In all three models $SU(2)$ gauge symmetry enhancement without
additional matter occurs at the boundary $T=U$. From the heterotic
point of  view,   this is the perturbative symmetry enhancement that
alway occurs in a  compactification on a circle of self-dual radius
$R=\sqrt{\alpha'}$. The additional massless states are momentum
and winding modes of the heterotic string around the $S^1$.

From the five-dimensional point of view, these additional massless
states appear pointlike. If one includes these extra modes, one
should therefore obtain an accurate description of the physics
near the boundary  $T=U$ in terms of an effective 5D field theory.
This field theory describes the coupling of altogether four vector
multiplets to 5D, $\mathcal{N}=2$ supergravity. In the following
section, which constitutes the main part of this paper, we will
derive this field theory.

\section{Gauged five-dimensional supergravity and singular
Calabi-Yau threefolds}
\setcounter{equation}{0}
From our discussion  in Section 2.3 it follows that the effective
field theory we are looking for is described by a  coupling of four
vector multiplets to 5D, $\mathcal{N}=2$ supergravity. However,
this coupling cannot be of the type described in Section 2.1,
because those theories were ``ungauged'' (that is, no field was
charged under any local gauge group). Instead, we now need  a
theory with a Yang-Mills-type $SU(2)$ gauge symmetry. The correct
framework is therefore what is commonly referred to as ``gauged''
supergravity. Gauged supergravity theories can be obtained
from their ungauged relatives by ``gauging'' appropriate
subgroups of the global (i.e., rigid) symmetry groups of the
latter. We briefly recall the most relevant aspects of this
procedure in Section 3.1, and then, in Sections 3.2 - 3.5, specialize
everything to the case we are interested in. For more details on
these theories the reader is referred to refs.
\cite{GST2,GZ1,CD,EGZ}.

\subsection{The gauging in general}

The ungauged Lagrangian (\ref{Lagrange1}) of Section 2.1 is invariant under a
global symmetry group of the form
\begin{displaymath}
SU(2)_R \times G.
\end{displaymath}
Here, $SU(2)_R$ denotes  the R-symmetry group of the underlying
Poincar\'{e} superalgebra. $SU(2)_R$ acts only on the index $i$ of
the fermions; all other fields, including the vector fields
$A_{\mu}^{I}$, are $SU(2)_R$ inert.   The group $G$, on the other
hand, is the invariance group of the cubic polynomial
$\mathcal{V}(h)$  (this group might very well be trivial,
depending on the polynomial). More precisely, $G$ is generated by
all (infinitesimal) linear transformations
\begin{eqnarray}
h^{I}&\longrightarrow& M^{I}_{J}h^{J}\label{htrafo}\\
A_{\mu}^{I}&\longrightarrow& M^{I}_{J}A_{\mu}^{J}\label{Atrafo}
\end{eqnarray}
that leave the tensor  $C_{IJK}$ (and therefore the cubic polynomial
$\mathcal{V}(h)$) invariant:
\begin{equation}\label{Cinv}
M^{I'}_{\,\,(I} C_{JK)I'}=0.
\end{equation}
Whenever such a non-trivial invariance group of $\mathcal{V}(h)$
exists,
it extends to a global symmetry group of the full Lagrangian, because the
latter is uniquely determined by $\mathcal{V}(h)$. In particular, all
$G$ transformations  act as isometries on the scalar manifold $\M$
(the converse is not true in general \cite{dWvP1}, though in most cases).
By definition, the $(n_V+1)$ vector fields $A_{\mu}^{I}$ transform in a
(not necessarily irreducible) $(n_V + 1)$-dimensional
representation of $G$.

If one wants to gauge a non-Abelian subgroup, $K$, of $SU(2)_R\times G$,
some of the vector fields $A_{\mu}^{I}$ have to transform in the adjoint
representation
of $K$.
Since all the vector fields are $SU(2)_R$ inert, such a
non-Abelian gauge group $K$ can only  be a subgroup of $G$ with the
additional constraint that
the $(n_V+1)$-dimensional representation of $G$ contains the adjoint
of $K$ as a subrepresentation. That is,  one needs a decomposition of the form
\be
\mathbf{(n_V+1)}_{G} = \mbox{adjoint}(K) \oplus \mbox{singlets}(K) \oplus
\mbox{non-singlets}(K).
\ee
As indicated, the $\mathbf{(n_V+1)}$ of $G$ might in general also
contain $K$-singlets as well as other non-singlet representations of $K$
in addition to the adjoint. Such additional non-singlets cause
technical complications
for the gauging of $K$ and  require the dualization of the
corresponding vector fields to ``self-dual'' tensor fields \cite{GZ1}.
Luckily, this case cannot occur in the  examples considered
in this paper,
and we can from now on assume a decomposition of the form
\be\label{decomp2}
\mathbf{(n_V+1)}_{G} = \mbox{adjoint}(K) \oplus \mbox{singlets}(K).
\ee
We split the embedding coordinates (and also the vector fields
$A_{\mu}^{I}$) accordingly,
\begin{equation}
h^I=(h^A,h^{\alpha}), \qquad
A_{\mu}^I=(A_{\mu}^A,A_{\mu}^{\alpha}),
\end{equation}
where $h^A$ (and $A_{\mu}^A$)  transform in the adjoint and
$h^{\alpha}$ (and $A_{\mu}^{\alpha}$) are singlets of $K$.

As mentioned above, $K\subset G$ acts on the scalar manifold $\M$
 via isometries.  The corresponding Killing vectors are given by \cite{GST2}
\be\label{Killing}
K^x_A = - \sqrt{ \frac{3}{2} } f^C_{AB} h_C h^{Bx},
\ee
where $f_{AB}^{C}$ denote the structure constants of $K$, and $h^{Bx} :=
g^{xy}h_y^B$  (see eq. (\ref{hIx})).

The gauging of $K$  then  proceeds in two steps \cite{GST2,GZ1}:
\begin{enumerate}
\item
Covariantization of all derivatives and field strengths with respect to $K$.
(For the Chern-Simons term this is slightly different \cite{GST2}, but
the details are irrelevant here.) This covariantization in general breaks
supersymmetry.
\item
Addition of further (gauge invariant) terms to the Lagrangian and to the
transformation rules in order to restore supersymmetry.
 For the case
at hand, one only needs to add the  Yukawa-like term \cite{GST2}
\begin{equation}\label{Yukawa}
e^{-1}\mathcal{L}_{\mscr{Yuk}}=
-\frac{i}{2}g{\bar{\lambda}}^{ix}\lambda_{i}^{y}K_{A[x} h^{A}_{y]}
\end{equation}
to the (covariantized) Lagrangian, while the (covariantized)
transformation rules do not receive any further corrections.
Here $g$ denotes the gauge coupling.
\end{enumerate}

Once $K$ is spontaneously broken, the term (\ref{Yukawa}) gives
rise to mass terms for some gaugini, indicating the presence of a
supersymmetric Higgs effect. Note that, in contrast to what
happens in four dimensions \cite{deWvPr84,ABCDFFM}, there  is no
scalar potential in  this kind of theories. This is consistent
with the fact that a massless vector multiplet in 5D  contains
only  one real scalar, which is eaten by the vector field once $K$
is broken. Thus, no mass term for scalar fields (and therefore no
potential) is needed. We will take a closer look at the Higgs
effect in Section 3.5.

\subsection{The gauging in our model}

After this general discussion, we now return to the
particular situation   we are  interested in.

The models described in Section 2.3 generically have two massless,
neutral vector multiplets, but at the bounary of the K\"ahler
cone two additional charged vector multiplets become massless and enhance
one of the $U(1)$s to the non-Abelian gauge group $SU(2)$. In order
to incorporate these two additional vector multiplets into a
complete field theoretical description, one therefore needs to find a
supergravity theory with $n_V=4$ vector multiplets in which
a subgroup $K=SU(2)$ of $G$ is gauged such that a Higgs effect
gives mass to two of the vector multiplets (except at those
points in moduli space where the full $SU(2)$ is unbroken).

The scalar manifold of such a theory is  four-dimensional
  and will be denoted
by  $\hat{\cal M}$. Likewise, we will put a hat on the corresponding
prepotential $\mathcal{V}$,
the $C_{IJK}$ and the $h^I$, i.e., we will write
\be
\hat{\cal V} = \hat{C}_{IJK} \hh^I \hh^J \hh^K, \qquad I,J,K=0,\ldots 4,
\ee
etc. in order to distinguish these quantities  from the analogous
quantities in the
ungauged
models of Section 2.3.

Our goal in this section is to determine the cubic polynomial
$\hat{\cal V}(\hat{h})$, which then uniquely specifies the corresponding
supergravity theory. $\hat{\cal V}(\hat{h})$ has
to fulfill the following minimal requirements:

\begin{enumerate}
\item
 $\hat{\cal V}(\hat{h})$ has to admit $K=SU(2)$ as an invariance group.
(Whether or not $K$ is possibly a genuine subgroup of a bigger invariance group
$G$ is not relevant for our purposes and will therefore not be
investigated.) Moreover, with respect
to $K=SU(2)$, the
five $\hat{h}^I$ have to decompose into the adjoint of $SU(2)$
plus two singlets  (cf. eq.   (\ref{decomp2})).
We will take  $\hat{h}^A$ $(A=1,2,3)$ to form the
adjoint and choose
$ \hat{h}^{\alpha}$ $(\alpha=0,4)$ to be the $SU(2)$ singlets.
\item
There is at least one point $c \in \hat{\cal M}$ where $SU(2)$ is
unbroken, i.e.,  this point  $c$ is  invariant under the action of the
$SU(2)$ isometry group.
(This requirements actually follows from the others \cite{EGZ}.
But for the problem
we are trying to solve, it is clear that we need this to be true,
so we impose it anyway because it will simplify the analysis.)
\item
The matrices $\stackrel{\circ}{a}_{IJ}$ and $g_{xy}$ should be
positive definite at $c$. This is necessary to have positive definite
kinetic terms at that point. From the supergravity point of view
the scalar manifold
is the maximal domain obtained by analytic continuation, such that
$\stackrel{\circ}{a}_{IJ}$ and $g_{xy}$ remain positive definite. In other
words, boundaries of the scalar manifold are loci where these
matrices degenerate or become singular. Note that on the boundary
of the K\"ahler cone $\stackrel{\circ}{a}_{IJ}$ and $g_{xy}$ can be regular
\cite{CKRRSW}.
Therefore the scalar manifold of M-theory has a different global
structure.
\end{enumerate}

In ref. \cite{EGZ}, all cubic polynomials that admit a compact
gaugeable invariance group (and which give rise to positive definite
$\stackrel{\circ}{a}_{IJ}$ and $g_{xy}$) were classified. Using that
classification, one immediately arrives at eq. (\ref{5N}) below as
the most general
solution to our constraints (i)-(iii). In order to keep our presentation
self-containted
we briefly explain how this result is  obtained.

Imposing $SU(2)$ invariance on $\hat{\cal V}$ implies
\bea
\hat{C}_{\alpha \beta A} &=& 0 \\
\hat{C}_{\alpha AB } &=& C_{\alpha} \delta_{AB} \\
\hat{C}_{ABC} &=& 0
\eea
while $\hat{C}_{\alpha \beta \gamma}$ is unconstrained. Here,
$C_{\alpha}$ are some undetermined coefficients.
To see this, recall  that    the $A$ index transforms in the adjoint, i.e.,
in the
$\mathbf{3}$ of
$SU(2)$, whereas $\alpha$ is a singlet index.
While there is a unique invariant in
$(\mathbf{3} \otimes \mathbf{3})_{\mscr{sym}}$, namely the
Cartan-Killing form (which can always be taken to be proportional to
$\delta_{AB}$ after an appropriate change of basis)
there are no invariants
in the $\mathbf{3}$ and in the symmetric part of $(\mathbf{3} \otimes
\mathbf{3} \otimes \mathbf{3})$.

Hence, group theory already restricts the polynomial to be
of the form
\begin{equation}\label{intermediate}
\hat{\mathcal{V}}(\hat{h})=
{\hat{C}}_{\alpha \beta \gamma} {\hat{h}}^{\alpha} {\hat{h}}^{\beta}
{\hat{h}}^{\gamma}
+3 C_{\alpha}{\hat{h}}^{\alpha} \delta_{AB}{\hat{h}}^{A}{\hat{h}}^{B}
\end{equation}

The  coefficients ${\hat{C}}_{\alpha \beta\gamma}$ and $C_{\alpha}$
can be fixed further by exploiting constraints (ii) and (iii).
According to constraint (ii), there is at least one $SU(2)$
invariant point, $c$, on $\hat{\M}$. An $SU(2)$ invariant point on $\hat{\M}$
has to have embedding coordinates ${\hat{h}}^{I}(c)=(c^0,0,0,0,c^4)$,
i.e., ${\hat{h}}^{A}(c)=0$. The coordinates $(c^0,c^4)$ define a
non-trivial direction in the $({\hat{h}}^0,{\hat{h}}^4)$ plane, because
${\hat{\mathcal{V}}}({\hat{h}}^{I}(c))=1$
implies that $c^0$ and $c^4$ cannot simultaneously be zero.
One can then always redefine ${\hat{h}}^0$ and ${\hat{h}}^4$ such that the new
coordinate values of the $SU(2)$ invariant point $c$ are given by
$({\hat{h}}^0(c),{\hat{h}}^4(c))=(1,0)$ (and of course still ${\hat{h}}^{A}(c)
=0$).

As $c$ is on $\hat{\M}$, we must have
\begin{displaymath}
{\hat{\mathcal{V}}}({\hat{h}}^{I}=(1,0,0,0,0))= \hat{C}_{000}=1
\end{displaymath}
so that in terms of the new ${\hat{h}}^0$ and ${\hat{h}}^4$ coordinates the
polynomial (\ref{intermediate}) becomes
\begin{equation}\label{intermediate2}
\hat{\mathcal{V}}(\hat{h})=(\hh^0)^3+a(\hh^0)^2 \hh^4 +b\hh^0(\hh^4)^2+
c(\hh^4)^3
+(d\hh^0+e\hh^4)\delta_{AB}\hh^{A}\hh^{B}
\end{equation}
with some coefficients $a,\ldots,e$.

The redefinition
\begin{eqnarray}
{\tilde{\hh}}^{0}&=&\hh^0+\frac{a\hh^4}{3} \\
{\tilde{\hh}}^{I}&=&\hh^I \qquad \textrm{for } I\neq 0
\end{eqnarray}
then removes the term quadratic in $\hh^{0}$ and changes the
coefficients $(b,c,d,e)$ to some values
$(\tilde{b},\tilde{c},\tilde{d},\tilde{e})$. Thus, after dropping
all the tildes again, the polynomial reads
\begin{equation}\label{intermediate3}
\hat{\mathcal{V}}(\hh)=(\hh^0)^3 +b\hh^0(\hh^4)^2+c(\hh^4)^3
+(d\hh^0+e\hh^4)\delta_{AB}\hh^{A}\hh^{B}.
\end{equation}

We will now make use of condition (iii). A necessary requirement
for the metric ${\stackrel{\circ}{a}}_{IJ}$ at $c$ to be positive
definite is that all its diagonal elements are positive at that
point. Using that $\hh^{I}(c)=(1,0,0,0,0)$, one finds for these
diagonal elements (cf. eqs. (\ref{aij}), (\ref{a0}))
\begin{eqnarray}
{\stackrel{\circ}{a}}_{00}(c)&=&1\\
{\stackrel{\circ}{a}}_{11}(c)&=&-\frac{2}{3}d={\stackrel{\circ}{a}}_{22}(c)
={\stackrel{\circ}{a}}_{33}(c)\\
{\stackrel{\circ}{a}}_{44}(c)&=&-\frac{2}{3}b.
\end{eqnarray}
Hence, $d$ and $b$ have to be negative, and a suitable rescaling
of $\hh^{A}$ and $\hh^{4}$ can be used to finally bring the polynomial
(\ref{intermediate3}) into the form
\begin{equation}\label{5N}
\hat{\mathcal{V}}(\hh)=(\hh^{0})^{3}-\frac{3}{2}\hh^{0}(\hh^{4})^{2}+\lambda
(\hh^{4})^{3}-\frac{3}{2}(\hh^{0}+\kappa \hh^{4})\delta_{AB}\hh^{A}\hh^{B}
\end{equation}
with some as yet arbitrary parameters $\kappa$ and $\lambda$.
The polynomial is now in the so-called canonical basis \cite{GST1,EGZ},
for which ${\stackrel{\circ}{a}}_{IJ}(c)=\delta_{IJ}$, so that
positivity of ${\stackrel{\circ}{a}}_{IJ}$ and $g_{xy}$ is manifest
in the vicinity of $c$.\\
\textbf{Remark 1:} Whenever $\lambda$ is negative, the
redefinition $\hh^4\rightarrow -\hh^4$
can be used to achieve
$\lambda \geq 0$, as we will assume from now on.\\
\textbf{Remark 2:} For precisely two pairs of values $(\kappa,\lambda)$,
the manifold $\hat{\M}$
becomes a symmetric space \cite{GST1,GST3,dWvP2}:
\begin{enumerate}
\item $(\kappa,\lambda)=(\sqrt{2},\frac{1}{\sqrt{2}})$
 $\Longrightarrow$     ${\cal M}=SO(1,1)\times SO(3,1)/SO(3)$.
\item $(\kappa,\lambda)=(-\frac{1}{\sqrt{2}},\frac{1}{\sqrt{2}})$
 $\Longrightarrow$    ${\cal M}=SO(4,1)/SO(4)$.
\end{enumerate}
\textbf{Remark 3:} $SU(2)$ invariant points on $\hat{\M}$
correspond to ${\hat{h}}^A=0$,
they thus form a codimension 3 hypersurface (i.e., a line) in the
four-dimensional scalar manifold $\hat{\cal M}$ given by $\hat{\mathcal{V}}
({\hat{h}}^I)
=1$.
On this line, the $SU(2)$ Killing vectors $K_{A}^{x}$ have to
vanish, which can also be verified   directly using the explicit expression
(\ref{Killing}) for the Killing vectors. This shows that the
Yukawa-type term (\ref{Yukawa}) vanishes there too, and the
gaugini (and with them the vector fields) have to be massless at
these points, confirming the naive expectation. We refer to Section
3.5 for a more detailed discussion of the Higgs mechanism.

The problem we seemingly have to solve now is the following:
can one embed
the theory defined by ${\cal V} = STU + \ft13 U^3$ into the
model defined by (\ref{5N})
such that the line $T=U$ on $\M$ coincides
with the line of $SU(2)$ invariant points on $\hat{\M}$?

Since we have to allow for
linear redefinitions of the embedding coordinates, this amounts to
expressing ${\hat{h}}^I$, $I=0,\ldots,4$  as linear combinations of $S,T,U$
such that
${\hat{h}}^A=0$ (i.e., $SU(2)$ invariance) is implied by $T=U$, and
the polynomials
are related by
\be
\hat{\cal V}|_{ {\cal M} \subset \hat{\cal M}}
= STU + \ft13 U^3.
\ee
This problem gives a coupled system of cubic equations
for the coefficients appearing in the linear relations between the two sets of
variables. With some
effort (see Appendix A)   one can convince oneself
that this system has {\bf no}
solution. Something is missing.

\subsection{Truncation versus integrating out}

The hidden assumption in the above reasoning is that it is sufficient
to set two of the five embedding coordinates to constant values in order
to eliminate the dynamics of the corresponding vector multiplets.
In field theory terms this means that we simply set the fields of these
 two vector multiplets to zero. But in a quantum theory it is not
sufficient to eliminate states as external lines, since they can
occur as internal lines and run in loops.

This is precisely what happens here. If one starts with   a theory with
four vector multiplets and Lagrangian $\hat{\cal L}$ and integrates
out two vector multiplets to get an effective Lagrangian ${\cal L}$
for two vector multiplets, then the two Lagrangians are related by
\be
{\cal L} =  \hat{\cal L}|_{\cal M} + \delta {\cal L}.
\ee
Here,  $ \hat{\cal L}|_{\cal M}$ is obtained by
restricting $\hat{\cal L}$ to the subspace ${\cal M}\subset \hat{\M}$
(which corresponds to simply \emph{truncating} out two vector multiplets),
 and
$\delta {\cal L}$ contains the effective interactions generated
by loops involving the two vector multiplets one integrates out.

Note that our problem is actually the reverse: we know ${\cal L}$,
which is given by the  polynomial $\mathcal{V}=STU+\frac{1}{3} U^{3}$,
and eventually want to reconstruct $\hat{\cal L}$. This is done by first
determining
 $\hat{\cal L}|_{\cal M}$ and then by fixing
the undetermined parameters $\kappa$ and $\lambda$ in
 $\hat{\cal L}$ such that restricting $\hat{\cal L}$ to $\M\subset \hat{\M}$
gives  $ \hat{\cal L}|_{\cal M}$.
(At this stage it is not clear whether this second step is
uniquely possible; it could be that the embedding of
$ \hat{\cal L}|_{\cal M}$ into  $\hat{\cal L}$
leaves room for an ambiguity in the parameters $\kappa$ and $\lambda$.)

To take the first step, we need to find $\delta {\cal L}$.
This problem is simplified by the fact that the whole Lagrangian
is determined by the coefficients $C_{IJK}$.  The only effect
of integrating out charged multiplets is a discrete change of these
coefficients. In order to trace this, it is sufficient to look
at the Chern-Simons term.

The precise form of the induced Chern-Simons term was computed in
\cite{EW96} for the case when a single charged hypermultiplet
is integrated out. In \cite{MS} this was generalized to the case
where both charged vector multiplets and charged hypermultiplets
are integrated out, with the charged hypermultiplets in arbitrary
representations of the gauge group. We will only need the case when
vector multiplets and hypermultiplets are in the same representation
(ultimately, the adjoint of $SU(2)$). Without loss of generality
we can parameterize the gauge fields such that the charged states
only couple to $A_{\mu}^{\star}$, where $\star$ is one particular
value in the range of $I$.

The result is most conveniently expressed in terms of the
corresponding prepotentials. If
${\cal V} = \hat{\cal V}|_{ {\cal M} } +
\delta {\cal V}$, then, in our conventions,
\be
\delta {\cal V} = \frac{1}{2} ( N_H - N_V ) (h^{\star})^3 \;,
\label{deltaV}
\ee
where $N_H$ and $N_V$ are the numbers of hypermultiplets and
vector multiplets charged under $A^{\star}_{\mu}$, and $h^{\star}$
denotes the corresponding embedding coordinate.
Compared to \cite{EW96,MV96,KMP} the above and various of the following
formulae differ by a factor $\ft16$. This is due to our different
normalization of the scalar fields.

\subsection{The embedding}

We now apply this general result to our case, where $N_H=0$, $N_V=2$
and $h^{\star} = h^1 = 6^{-1/3} (T-U)$ and obtain
\be\label{STUneu}
\hat{\cal V}|_{\cal M} = STU + \ft13 U^3
+ \ft16 (T-U)^3 \;.
\ee

As indicated by our notation, this polynomial should now be the one  that
one obtains from (\ref{5N}) by a suitable restriction to the three
variables $S,T,U$. More precisely, expressing $\hh^0,\ldots,\hh^4$
as  appropriate linear combinations of $S,T,U$, the polynomial
$\hat{\mathcal{V}}$ (eq. (\ref{5N})) should reduce to (\ref{STUneu}).
By ``appropriate''
linear combinations, we mean that $T=U$ should imply $SU(2)$ invariance, i.e.,
 $\hh^1=\hh^2=\hh^3=0$. This results in a system of ten coupled cubic
 equations
for the coefficients appearing in these linear combinations (see Appendix
A). Remarkably, a solution of this  system exists \emph{if and only if}
the coefficient in front of  $(T-U)^3$ in (\ref{STUneu}) is really
$\frac{1}{6}$, i.e., exactly as the integrating out procedure suggests.
For all other values of this coefficient, no solution exists.
This reconfirms the importance of properly taking into account the one-loop
threshold effects and is an important consistency check.

Taking now the right coefficient $\frac{1}{6}$, one further finds
 (see Appendix A) that $\hh^0,\ldots,\hh^4$ can be consistently expressed
 in terms of $S,T,U$
if and only if the coefficients $\kappa$ and $\lambda$ satisfy the constraints
\be
\lambda = \frac{1}{\sqrt{2}} \;,\;\;\;
\kappa > - \lambda \;, \;\;\; \kappa \not= 2 \lambda \;.
\ee
Note that the two excluded boundary values of $\kappa$,
namely $\kappa = -  \frac{1}{\sqrt{2}}$ and $\kappa = \sqrt{2}$
are precisely those values for which $\hat{\cal M}$ would have been
 a symmetric
space.

It now seems that our embedding problem does have a consistent solution,
but that this solution is not unique. In fact, there seems to be a
one-parameter family of Lagrangians $\hat{\cal L}$ parameterized by the
allowed values of
$\kappa$ in the underlying cubic polynomial $\hat{\mathcal{V}}$.
Surprisingly, this is not true.
As is shown in Appendix B, every pair of admissible values
$\kappa_1$, $\kappa_2$ can be transformed into each other
by a linear redefinition of the  $\hat{h}^I$. Therefore all these manifolds
are physically equivalent, and the embedding is unique.
The unambiguous construction of this supergravity theory from the underlying
Calabi-Yau data is the main result of this paper.

For the sake of completeness, we note the
 explicit relations between $S,T,U$ and $\hat{h}^I$:
\bea
\hat{h}^0 &=& \frac{1}{3 G^2} \left(
\frac{2U-T}{3} + S \right) + \frac{G}{3} (T+U) \label{hhat0}\\
(\hat{h}^1)^2 + (\hat{h}^2)^2 + (\hat{h}^3)^2 &=& \left(
\frac{G^2}{6} - \frac{2}{9G} \right) (T-U)^2 \\
\hat{h}^4 &=& \frac{\sqrt{2}}{3} \left(
\frac{ 2U-T }{ 3 G^2} + \frac{ S }{G^2} - \frac{G}{2} (T+U) \right)
\label{hhat4}\;,
\eea
where
\be
G^3 = - \frac{4}{3} \left(\frac{1  + \sqrt{2} \kappa }{ 2 - \sqrt{2} \kappa}
\right).
\ee
Observe that $(T-U)^2$ is related to the gauge invariant combination
$(\hat{h}^1)^2 + (\hat{h}^2)^2 + (\hat{h}^3)^2$.
By a gauge transformation we can set $\hat{h}^3 \propto T-U$, and
then the truncation from $\hat{\cal M}$ to ${\cal M}$ is given
by $\hat{h}^1 = \hat{h}^2 = 0$.

\subsection{The Higgs effect}

The purpose of this subsection is to explicitly  recover the
expected Higgs effect   in the $SU(2)$ gauged supergravity theory
based on $\hat{\mathcal{L}}$. \footnote{In order to simplify the
notation, we will not put hats on the fields appearing in
$\hat{\mathcal{L}}$.}

Due to the absence of a scalar potential, any vev $\langle
\phi^{x}\rangle $ of the scalar fields $\phi^{x}$ gives rise to a
Minkowski ground state of the theory, which can be easily verified
to  preserve the full $\mathcal{N}=2$ supersymmetry \cite{GST2}
(provided that $\langle A_{\mu}^{I}\rangle= \langle \lambda^{ix}
\rangle =\langle \psi_{\mu}^{i} \rangle =0$).

Any vev $\langle \phi^{x}\rangle $ of  $\phi^{x}$ also determines
a vev $\langle  \hh^{I} \rangle$ of the corresponding embedding
coordinates.  The vacua with $\langle \hh^{A}\rangle =0$, $A=1,2,3$,
leave the full
$SU(2)$ gauge symmetry unbroken, whereas a non-vanishing value for
at least one of the three $\langle \hh^{A} \rangle$ spontaneously
breaks  $SU(2)$ to $U(1)$. In the latter case, the Calabi-Yau
picture requires two vector multiplets to become massive BPS
vector multiplets, in which the former scalar fields $\phi$
 contribute the longitudinal modes of the massive `W-bosons'.

If we choose $\langle \hh^{3} \rangle$ to be the only
non-vanishing vev of $\hh^{A}$, the vector fields $A^{1,2}_{\mu}$
and the gaugini $\lambda^{i 1,2}$ should therefore  acquire a mass
proportional to $\langle \hh^{3} \rangle$  (see eq.
(\ref{BPSmass})).

To see how this happens, consider
\begin{eqnarray}
A_{\mu}^{0}&:=&\langle\hh_I \rangle A_{\mu}^{I}, \qquad
F_{\mu\nu}^{0}=\langle\hh_I \rangle F_{\mu\nu}^{I}\label{F0}\\
A_{\mu}^{x}&:=&\langle\hh_I^x \rangle A_{\mu}^{I}, \qquad
F_{\mu\nu}^{x}=\langle \hh_I^x \rangle F_{\mu\nu}^{I},\label{Fx}
\end{eqnarray}
where $\hh_{I}$ is as in eq. (\ref{hdual}), and $\hh_I^x$ is
defined by \cite{GST1}
\begin{equation}
\hh_{I}^{x}:=\sqrt{\frac{3}{2}}g^{xy} \frac{\partial
\hh_I}{\partial \phi^y}.
\end{equation}
In a given vacuum,  the  vector fields $A_{\mu}^{x}$ are the
superpartners of $\lambda^{ix}$ and $\phi^{x}$, whereas
$A_{\mu}^{0}$ is the graviphoton  (i.e.,  $A_{\mu}^{x}$ and
$A_{\mu}^{0}$ appear in the supersymmetry transformations of,
respectively,  the gaugini and the gravitini \cite{GST2}).

In order to see which of these vector fields acquires a mass, one
uses the identity \cite{GST1}
\begin{equation}
\stackrel{\circ}{a}_{IJ} =\hh_I \hh_J +g_{xy} \hh_{I}^x \hh_{J}^y
\end{equation}
to rewrite the kinetic term of the vector fields in a given
vacuum:
\begin{equation}
e^{-1} \hat{\cal L}_{\mscr{vec}} =
-\frac{1}{4}\langle \stackrel{\circ}{a}_{IJ}\rangle F_{\mu\nu}^{I}
F^{\mu\nu J}=
-\frac{1}{4} F_{\mu\nu}^{0} F^{\mu\nu 0} -\frac{1}{4} \langle g_{xy} \rangle
F_{\mu\nu}^{x} F^{\mu\nu y}.
\end{equation}
Mass terms for these vector fields can only come from the kinetic
term of the scalar fields,
\be \label{scalarkinetic}
e^{-1} \hat{\cal
L}_{\mscr{scalar}} = - \ft12 g_{xy} {\cal D}_{\mu} {\phi}^x {\cal
D}^{\mu} {\phi}^y, \ee where ${\cal D}_{\mu}$ denotes the  $SU(2)$
covariant derivative of the scalar fields \cite{GST2},
 \be {\cal D}_{\mu}
{\phi}^x = \der_{\mu} {\phi}^x + g A^A_{\mu} K_A^x. \ee Using some
identities of very special geometry \cite{GST1,GST2}, one can show
that the term quadratic in the vector fields in
(\ref{scalarkinetic}) can be brought to the form
\begin{equation}
\frac{1}{2} (A_{\mu}^{x} A^{\mu y}) (g W_{xz}) (g W^z_{\,\,\,y}),
\end{equation}
where $W_{xy}:=h_{[x}^{A}K_{Ay]}$.

Given  the kinetic term for the gaugini  (see   \cite{GST2})
\begin{equation}
-\frac{1}{2}({\bar{\lambda}}^{ix} \Gamma^{\mu} \nabla_{\mu}
\lambda_{i}^{y}) g_{xy}
\end{equation}
and the Yukawa term (\ref{Yukawa}) required by supersymmetry,
\begin{equation}
e^{-1}\hat{\mathcal{L}}_{\mscr{Yuk}}=\frac{i}{2}({\bar{\lambda}}^{ix}
\lambda_{i}^{y}) (gW_{xy}),
\end{equation}
it is now easy to see  that, in a given vacuum with a given
vev $\langle W_{xy} \rangle$, $A_\mu^{x}$ and $\lambda^{ix}$
automatically have the same mass, and that the
graviphoton $A_{\mu}^{0}$ remains massless, as it should. These
observations are of course automatic consequences of the unbroken
$\mathcal{N}=2$ supersymmetry of these vacua.

What remains to be verified  is whether a non-vanishing vev $\langle
\hh^{3} \rangle $ really introduces a mass proportional to
$\langle \hh^{3} \rangle $ for $A_{\mu}^{1,2}$ and $\lambda^{i
1,2}$, and whether all other fields remain massless in such a
vacuum. According to what we have said above, it is sufficient to
look at the masses of the fermions $\lambda^{ix}$.

To this end, we choose the `special coordinates'
\be
{\phi}^x = \frac{\hat{h}^x}{\hat{h}^0} \;,\;\;\;x=1, \ldots , 4 \;
\ee
to parameterize the scalar manifold $\hat{\M}$.
This choice allows one to explicitly solve  $\hat{h}^I$
as functions of the ${\phi}^x$ using the constraint
$\hat{\cal V} =1$.

For the Yukawa term, one then obtains
\be
e^{-1} \hat{\cal L}_{\mscr{Yuk}}
= \frac{i}{2} \overline{\lambda}^{iA} \lambda_i^B
\left[  \left( \ft{3}{2} \right)^{3/2}  g
(1 + \kappa {\phi}^4)
(\hat{h}^0)^3 \right] \epsilon_{ABC} \hat{h}^C \;.
\ee

We will now allow $\langle \hh^{3} \rangle$ and  $\langle \hh^{4} \rangle$
to be non-zero, whereas $\langle \hh^{1} \rangle$ and
$ \langle \hh^{2} \rangle$ are assumed to vanish.

The  vev of the mass matrix $W_{xy}$ is then
\begin{equation}
\langle W_{xy} \rangle =\left(\begin{array}{cccc}
0 & a & 0 & 0\\
-a& 0&0&0\\
0&0&0&0\\
0&0&0&0
\end{array}\right),
\end{equation}
where $a=(\frac{3}{2})^{\frac{3}{2}}  (1+\kappa \langle \phi^4 \rangle )
\langle \hh^0 \rangle^3 \langle \hh^3 \rangle$.

In order to really extract the physical masses from this matrix, one has
to make sure that one works in a field basis with canonically normalized
kinetic terms.

In our class of vacua, where $\langle \hh^1 \rangle =\langle \hh^2 \rangle
=0$,  the vev  of the metric $g_{xy}$ turns out to be of the form
\begin{equation}
\langle g_{xy} \rangle =\left(\begin{array}{cccc}
b & 0 & 0 & 0\\
0&  b    &0&0\\
0&0&\langle g_{33} \rangle &\langle g_{34}\rangle\\
0&0&\langle g_{43} \rangle &\langle g_{44} \rangle
\end{array}\right)
\end{equation}
with $b:=\frac{3}{2} (1+\kappa \langle \phi^4\rangle)\langle \hh^{0}\rangle^3$

In order to obtain canonically normalized $\lambda^{ix}$,
one has to redefine the $\lambda^{ix}\rightarrow{\lambda'}^{ix} $ such that
${\bar{\lambda}}^{ix}\lambda^{y}_i\langle g_{xy} \rangle   =
{\bar{\lambda'}}^{ix}{\lambda'}^{y}_i
\delta_{xy}$.
Due to the peculiar form of $\langle g_{xy} \rangle$,
${\lambda'}^{\, i3}$ and ${\lambda'}^{\, i4}$ will be some linear
combinations of
${\lambda}^{i3}$ and ${\lambda}^{i4}$, whereas ${\lambda'}^{\, i1}$ and
${\lambda'}^{\, i2}$ are obtained by a mere rescaling,
\be
{\lambda'}^{\, i1,2}=b^{\frac{1}{2}} \lambda^{i1,2}.
\ee
In terms of these new, canonically normalized
 gaugini (whose primes we will omit from now on),
the Yukawa term becomes
\be \hat{\cal L}_{\mscr{Yuk}}^{\mscr{can}} =
\frac{i}{2} ( \overline{\lambda}^{i1} \lambda^2_i -
\overline{\lambda}^{i2} \lambda^1_i) Q  \langle \hat{h}^3
\rangle\ee with $Q = \sqrt{\ft32} g$. This shows   that
only the gaugini $\lambda^{i1,2}$ (and with them the vector fields
$A_{\mu}^{1,2}$) acquire a mass, and that this mass is indeed
proportional to $\langle \hat{h}^3 \rangle $, \be
m_{\lambda^{1,2}} = Q \langle \hat{h}^3 \rangle \;. \ee

Since the
charge of a BPS M2-brane is proportional to its tension, the
$SU(2)$ gauge coupling $g$ must be equal to $T_{(2)}$, up to a
constant which we do not attempt to compute here.

\section{Elementary transformations, reflected cones and the Weyl twist}
\setcounter{equation}{0}

In the extended scalar manifold $\hat{{\cal M}}$ there is no obstruction
in going to negative values of $\hat{h}^3 \propto T-U$, whereas in the
original model $T=U$ is the boundary of the K\"ahler cone, and a
continuation to negative $T-U$ seems to be impossible because the
Calabi-Yau space becomes singular at $T=U$. How can these two
facts be reconciled?

As in many similar cases, it makes sense to continue
the theory beyond the boundary of the K\"ahler cone. In general such
continuations involve a change of topology. The basic mechanism is that
the singular points of the singular manifold $\hat{X}$ can be resolved
in two different ways, thus giving rise to two different families
$X, \tilde{X}$ of smooth Calabi-Yau spaces. As will become more and more
clear during this section, the singular space $\hat{X}$ has a close
relation to the gauged Lagrangian $\hat{\cal L}$. This motivates our
notation.

The K\"ahler cones of the two families $X,\tilde{X}$ can be glued
together along the face corresponding to the singular manifold
$\hat{X}$. One example, where $X$ and $\tilde{X}$ have different
topology (triple intersection numbers) but are still birationally
equivalent are flop transitions \cite{EW96}. The case we are
considering, on the other hand,  corresponds to an elementary
transformation. In this case the families $X,\tilde{X}$ are
biholomorphically equivalent. It will turn out that the symmetry
relating $X$ to $\tilde{X}$ is nothing but an $SU(2)$ gauge
transformation.

In Section \ref{general} we will review some general facts about topological
phase transitions in Calabi-Yau threefolds, then, in Section \ref{ours}
we consider the elementary
transformation occuring in our models explicitly. In a little digression we
discuss the relevance of elementary transformations and reflected cones
for space-time geometries where $SU(2)$ gauge symmetry enhancement
occurs at special points in space-time.
Finally, we show in Section \ref{Weyltwist}
how the elementary transformation is realized in the $SU(2)$-gauged
effective supergravity that includes the additional light modes.

\subsection{General discussion of elementary transformations \label{general}}

This subsection is a short review of material from \cite{EW96,KMP}.

Let $h^I >0$ parameterize the K\"ahler cone of a Calabi-Yau
threefold $X$ and let $h^{\star} = 0$ be the boundary where the
two-cycle $C^{\star}$ collapses. ($\star$ is just one of the possible
values of $I$.)
Thus at the boundary $h^{\star}=0$ the volume of $C^{\star}$,
\be
\textrm{vol}(C^{\star})= \int_{C^{\star}} J=h^{\star},
\ee
vanishes and it becomes negative when one continues naively
to negative $h^{\star}$. Correspondingly, an
M2-brane wrapped on $C^{\star}$ seems to get a negative mass
\be
M = T_{(2)} h^{\star}
\ee
for negative $h^{\star}$.

But, as already explained above, the range of negative $h^{\star}$
actually corresponds to a different Calabi-Yau space $\tilde{X}$,
which is obtained by resolving the singularities of $\hat{X}$ in a
different way, and which in general has a topology different from the one
of $X$.

In the most general topological transitions the Hodge numbers,
and with them   the numbers of neutral vector and
hypermultiplets \cite{Str95,KM96,KMP}, change. This requires the presence
of at least two   adjoint hypermultiplets (or of
hypermultiplets in other representations of the gauge group) at
the singularity, so that the scalar potential has both a Coulomb
and a Higgs branch. In our case these necessary conditions are not
satisfied, and the Hodge numbers of $X$ and $\tilde{X}$ are the
same. Even if the Hodge numbers do not change, the manifolds $X$
and $\tilde{X}$ might still have different topology. If the triple
intersection numbers are different and are not related by a basis
transformation of the K\"ahler cone, then $X$ and $\tilde{X}$ have
different topology. But if the intersection numbers are
equivalent, then $X$ and $\tilde{X}$ have the same topology, and
the transformation relating them is an isomorphism (in fact a
gauge transformation as we will see below). In both cases one can
glue together the K\"ahler cones of $X$ and $\tilde{X}$ along the
face $h^{\star} = 0$. Within the K\"ahler cone of $\tilde{X}$ the
cycle $C^{\star}$ is replaced by $\tilde{C}^{\star}=-C^{\star}$,
which has positive volume $\tilde{h}^{\star} = - h^{\star}$ when
measured with the K\"ahler form $\tilde{J}$ of $\tilde{X}$. When
one goes from positive to negative $h^{\star}$, states obtained
from wrapping an M2-brane on $C^{\star}$ become massless and then
get a positive mass again.

The explicit transformation relating $X$ to $\tilde{X}$ depends
on the details of the degeneracy at $h^{\star} =0$. For
proper transformations (flops) and elementary transformations it can
be computed using methods from algebraic geometry. Alternatively,
one can analyze the additional massless modes present at
$h^{\star}=0$ and compute the change induced in the prepotential
by integrating them out \cite{EW96,MV96}. The result of this depends
on whether $h^{\star} > 0$ or $h^{\star} < 0$. These two cases differ
in the sign of the mass of the charged particles running in the loops
(where mass is defined from the point of view of manifold $X$, i.e.,
it becomes negative for $h^{\star} < 0$). Using the notation of
Section 3, integrating out charged states at $h^{\star} >0$
gives
\be
{\cal V} = \hat{\cal V} |_{\cal M} + \delta {\cal V},
\label{Vplus}
\ee
whereas the result for $h^{\star}<0$ is
\be
\tilde{\cal V} = \hat{\cal V} |_{\cal M} - \delta {\cal V}.
\label{Vminus}
\ee
Using (\ref{deltaV})
one finds the following discontinuity at $h^{\star} = 0$:
\be
\tilde{\cal V} - {\cal V} = - 2 \delta {\cal V}
= (N_V - N_H) (h^{\star})^3 \;.
\ee
This discontinuity is a five-dimensional one-loop threshold
effect \cite{AFT,EW96,MV96},
analogous to the well-known logarithmic singularity of the prepotential
in four dimensions \cite{CDFP,Str95}.
Since in five dimensions the whole vector multiplet sector of the
effective theory of uncharged massless modes is determined by the cubic
prepotential, a discrete jump of its coefficients (the triple intersection
numbers) is the only possible threshold effect.

This field theory calculation can be compared to the formulae
for proper and elementary transformations
obtained in algebraic geometry:
\begin{enumerate}
\item
Take the case $N_H=1$, $N_V=0$ of a single massless hypermultiplet.
In this case one obtains
\be
\tilde{\cal V} - {\cal V} = - (h^{\star})^3
\ee
which is the correct formula for a flop transition, where $X$ and
$\tilde{X}$ are related by a proper transformation \cite{EW96}. More
generally,
there are transitions where $N_H$ curves are flopped, and then
$\tilde{\cal V} - {\cal V} = - N_H (h^{\star})^3$.
\item
The case $N_H = 2g$, $N_V=2$ of $SU(2)$ gauge symmetry enhancement
with $g$ adjoint hypermultiplets. Geometrically this is realized by
the collapse of a divisor $E$ into a genus $g$ curve of $A_1$ singularities.
The supergravity formula gives
\be
\tilde{\cal V} - {\cal V} = (2 - 2g)  (h^{\star})^3
\ee
which coincides with the geometrical formula for an elementary
transformation relating $X$ and $\tilde{X}$ \cite{KMP}.
\end{enumerate}

Whereas proper transformations relate manifolds with
different topology, $X$ and $\tilde{X}$ are isomorphic
(biholomorphically equivalent) for elementary transformations.
This is obvious since the elementary transformation acts on the
K\"ahler cone by a reflection, which could be reinterpreted as a change
of basis in $H_2(X,\mathbb{Z})$. We will see this explicitly
in Section \ref{ours}, and in Section \ref{Weyltwist}
we will show that the basis transformation is an $SU(2)$
gauge transformation. In this case the K\"ahler cone of $\tilde{X}$ is
called the reflected cone. One might wonder whether
gluing in the reflected K\"ahler cone has any use, because the
original K\"ahler cone already covers all inequivalent manifolds once.
It turns out that working with a doubled range of variables is nevertheless
useful for several reasons:

\begin{enumerate}
\item
Working with the reflected cone is useful for studying solutions
which dynamically run into $SU(2)$ gauge symmetry enhancement.
This will be briefly explained in the next subsection and covered in detail
in \cite{TM1}.
\item
It is instructive to see what the elementary transformation is in
terms of gauged supergravity.
As we will see in
Section \ref{Weyltwist}, this makes use of the reflected K\"ahler cone.
\item
If the genus of the curve of $A_1$ singularities is $g>0$, then not only
charged vector multiplets but also charged hypermultiplets become
massless. When taking into account deformations of the complex structure,
which are hypermultiplet moduli, one finds that the elementary transformation
acts on both kinds of multiplets in such a way that the
factorization of the moduli space into vector multiplet moduli
and hypermultiplet moduli breaks down\cite{KMP}. The point of
enhanced symmetry is an orbifold point of the combined moduli space and
one needs to introduce the reflected cone to describe the moduli
space properly.
If at least two adjoint hypermultiplets are present,
then there exist extremal transitions which change the Hodge numbers of $X$
\cite{KM96,KMP,MV96}. This corresponds to a scalar potential which possesses
both a Coulomb and a Higgs branch.

\end{enumerate}

\subsection{Application to our models and consequences for space-time
geometries \label{ours}}

The case we have been studying in this paper is $N_H= 2g=0$, $N_V=2$.
Switching from the variables $h^I$, which are adapted to the K\"ahler cone,
to the variables $S,T,U$ and using the general formulae
(\ref{Vplus}) and (\ref{Vminus}),
we find the following prepotentials associated
to $X$ and $\tilde{X}$:
\bea
{\cal V} = STU + \ft13 U^3 & & \mbox{   for   }T>U \\
\tilde{\cal V} = STU + \ft13 U^3 + \ft13 (T-U)^3
 & & \mbox{   for   } T<U
\eea Note that one could make a basis transformation and bring the
prepotential for $T<U$ to the form $\tilde{\cal V} = STU + \ft13
T^3$ \cite{AFT}. But when working with the combined cones of $X$
and $\tilde{X}$ one has to apply the same change of basis to
${\cal V}$, which then no longer is of the form $STU + \ft13 U^3$.
This has important consequences for space-time geometries with
rolling moduli. When considering backgrounds where $h^{\star}$
starts at a positive value and dynamically evolves to zero, one
has to study whether and how such solutions can be continued to
$h^{\star} <0$. This problem was analyzed for domain walls in the
model with $X = X_{1}$ in \cite{KalMohShm}. Since both the metric
$g_{xy}$ on the moduli space and the space-time metric describing
the domain wall were found to be smooth, it was expected that one
could continue the solution using the method introduced in
\cite{GMMS}. But in \cite{KalMohShm} the prepotential used to
define the continuation was ${\cal V} = STU + \ft13 U^3$ for $T>U$
and $\tilde{\cal V} = STU + \ft13 T^3$ for $T<U$. Using this rule
for the continuation, it was found that the space-time Riemann
tensor of the domain wall solution has  $\delta$-function
singularities at the space-time points at which $T=U$. In other
words, one would have to introduce a localized source of
stress-energy at the points in space where $T=U$.

This is different when one uses the correct continuation
of the prepotential which we obtained
above by properly applying an elementary transformation. It
can be shown that there is no such $\delta$-function singularity, but instead
one finds that the space-time metric is $C^2$,
precisely as in flop transitions occuring dynamically in black hole
geometries \cite{CKRRSW,GMMS} and domain walls \cite{GSS}. In fact one can
show that continuing black hole, black string or domain wall solutions
through an $SU(2)$ boundary is equivalent to reflecting it at the
boundary by a gauge transformation. Using this one can prove that for
the models based on the Calabi-Yau threefolds $X_{0}, X_{1}, X_{2}$
no space-time singularities can occur as long as the moduli roll within
the extended K\"ahler cone. In particular, all apparent
space-time singularities found in supergravity are not really present in
M-theory, because the $SU(2)$ enhancement modifies the evolution of the
solution. This explains how
the new mechanism for excising space-time
singularities proposed in \cite{KalMohShm} works.
We refer to a future publication for the details \cite{TM1}.

We now turn to an explicit description of how $\tilde{X}$ is
obtained from $X$. In the next section we will show that this
transformation is an $SU(2)$ gauge transformation.

To construct $\tilde{X}$ from ${X}$, we need to specify how
the prepotential (triple-intersection numbers) and the
boundaries of the K\"ahler cones are related. Let $C^I$ and
$D_I$ again be the standard generators of $H_2(X,\mathbb{Z})$
and $H_4(X,\mathbb{Z})$.

There is a family of holomorphic curves $\Gamma$ which sweeps out
a divisor $E$. The elementary transformation acts by \cite{KMP}
\be D_I \longrightarrow \tilde{D}_I = D_I + (D_I \cdot \Gamma) E.
\label{elementary} \ee
Since $\Gamma \cdot E = -2$ \cite{KMP} one
has $E \rightarrow - E$, showing that the elementary
transformation acts as a reflection. Therefore the K\"ahler cone
of $\tilde{X}$ is called the reflected cone. The collection of all
cones obtained by reflections is called the reflected movable cone
\cite{KMP}.

One can now compute $\tilde{C}_{IJK}$, and the
new dual two-cycles $\tilde{C}^I$ are determined from
$\tilde{C}^I \cdot \tilde{D}_J = \delta^I_J$.

In our models the $SU(2)$ enhancement occurs for
$h^1=0$. This means that the curves $\Gamma$ are in the homology
class $C^1$. The dual four-cycles $D_I$ are determined by the
dual embedding coordinates $h_I = C_{IJK} h^J h^K$. We have to
identify a four-cycle which collapses to a two-cycle for
$h^1=0$. This identifies the homology class of the divisor $E$.
Since the intersection numbers enter, this cycle is different
for the three models $X_0, X_1, X_2$. To be a bit more explicit,
we consider the
model $X_1$.
There we have
\bea
h_0 &=& 8 (h^0)^2 + 6 h^0 h^1 + (h^1)^2 + 4 h^0 h^2 + 2 h^1 h^2 \\
h_1 &=& 3 (h^0)^2 + 2 h^0 h^1 + 2 h^0 h^2 \\
h_2 &=& 2 (h^0)^2 + 2 h^0 h^1.
\eea
There is a unique linear combination (up to an overall constant)
which vanishes for $h^1=0$:
\be
h = h_0 - 2 h_1 - h_2 = h^1 ( h^1 + h^2).
\ee
Note that $h$ goes to zero $\sim h^1$, indicating that the
four-cycle collapses to a two-cycle rather than to a zero-cycle.
Thus we have in homology
\be
E = D_0 - 2 D_1 - D_2.
\ee
Observe that $C^1 \cdot E = -2$ as it must. Now applying formula
(\ref{elementary})
for an elementary transformation, one obtains the new four-cycles
\be
\tilde{D}_0 = D_0 \;,\;\;\;
\tilde{D}_1 = D_0 - D_1 - D_2 \;,\;\;\;
\tilde{D}_2 = D_2,
\ee
and the new dual two-cycles are
\be
\tilde{C}^0 = C^0 + C^1 \;,\;\;\;
\tilde{C}^1 = - C^1 \;,\;\;\;
\tilde{C}^2 = - C^1 + C^2 \;.
\ee
Now we can read off the transformation of the $h^I$:
\be
h^0  \rightarrow  h^0 + h^1 \;,\;\;\;
h^1 \rightarrow - h^1 \;,\;\;\;
h^2  \rightarrow -h^1 + h^2 \;.
\label{TransX}
\ee
Converting this to $S,T,U$ we find
\be\label{Strafo}
T \leftrightarrow U \; \;\;\; S \rightarrow S + U - T.
\label{TransSTU}
\ee

\subsection{The Weyl twist \label{Weyltwist}}

Given that the elementary transformation acts as a $\mathbb{Z}_2$
reflection, the natural explanation in terms of the gauge theory is
to identify it with the Weyl twist of $SU(2)$ \cite{KMP,MV96}.

Recall that the Weyl group of a simple Lie group $G$ consists of
all inner automorphisms that leave the maximal torus invariant
modulo those which do so pointwise:
\be
W(G) = N(T) / T,
\ee
where $W(G)$ is the Weyl group of $G$, and $N(T)$ is the normalizer of the
maximal torus $T$. The Weyl group of $SU(2)$ is isomorphic to $\mathbb{Z}_2$,
its generator is called the Weyl twist.

In our $SU(2)$ gauged supergravity theory we have fields $\hat{h}^A$ in the
adjoint and singlets $\hat{h}^0,\hat{h}^4$. The singlets are of course
invariant. Taking the maximal torus to be generated by $\hat{h}^3$,
the Weyl twist acts by $\hat{h}^3 \rightarrow - \hat{h}^3$, while
all other fields are invariant. Now we make use of the unique
solution of the embedding problem we found in Section 3.
Using the explicit
relations (\ref{hhat0}) - (\ref{hhat4}) between the variables $\hat{h}^I$
and the variables $S,T,U$ we immediately find that
$S,T,U$ precisely transform as in (\ref{TransSTU}).
This shows that the Weyl twist operates as an elementary transformation
on the Calabi-Yau spaces.

It is also instructive to consider the behaviour of the
prepotentials ${\cal V}, \hat{\cal V}, \tilde{\cal V}$ under the
Weyl twist.
By construction, the prepotential $\hat{\cal V}$ is invariant.
This is still true after projecting to ${\cal M}$, i.e.,
\be
\hat{\cal V}|_{\cal M} = STU + \ft13 U^3 + \ft16 (T-U)^3
\ee
is invariant under (\ref{TransSTU}). The prepotentials
${\cal V}$ and $\tilde{\cal V}$ that we obtained by integrating out
the charged vector multiplets for $T-U >0$ and $T-U <0$ are not
invariant but are precisely mapped to one another under
(\ref{TransSTU}). This shows explicitly that the models defined by
compactification on $X$ and $\tilde{X}$ are related by an $SU(2)$
gauge transformation.

It is interesting to ask what interpretation the full prepotential
$\hat{\cal V}$ might have in terms of Calabi-Yau geometry. Its coefficients
are not intersection numbers of either $X$ or $\tilde{X}$, but
$\hat{\cal V}|_{\cal M}$ is obtained as a `superposition' or
`orbit sum':
\be
\hat{\cal V}|_{\cal M} = \ft12 ( {\cal V} + \tilde{\cal V} )
\ee
while the extension to the full $\hat{\cal V}$ seems to be
dictated by gauge symmetry (this is at least what we realized
a posteriori above).

The above relation is true more
generally for integrating out $N_H$ hypermultiplets
and $N_V$ vector multiplets, even though in those cases the
$\mathbb{Z}_2$ need not be a symmetry transformation. In particular,
it is not a symmetry transformtion
for flops. The natural conjecture is that
$\hat{\cal V}$ plays a r\^{o}le in the intersection theory of
singular Calabi-Yau spaces. For example the coefficients
$\hat{C}_{IJK}$ $I,J,K = 0,\ldots,4$
could be used to construct a generalization of the
(co)homology ring defined by $C_{IJK}$, $I,J,K = 0,1,2$.
Gauge theory and gauged supergravity
might provide the tool to describe singular Calabi-Yau spaces as
regular objects, presumably by adding extra data (corresponding to
wrapped branes) located at the vanishing cycles.

The transformation (\ref{TransSTU}) is natural from the point
of view of the dual heterotic string. There, the transformation
$T \leftrightarrow U$ corresponds to inverting the radius
$R \rightarrow \ft{\alpha'}R$ \cite{AFT}.
At the fixed point $T=U \Leftrightarrow R=\sqrt{\alpha'}$ of this
transformation the Abelian Kaluza Klein gauge symmetry $U(1)$ is
enhanced to $SU(2)$. This is one example of the interpretation
of the T-duality group as a discrete gauge symmetry. For toroidal 
compactifications
it has been shown \cite{GPR}
that the full T-duality is generated by Weyl twists
of $SU(2)$ gauge groups that are un-Higgsed at special loci of the
moduli space. The fact that $S$, the heterotic dilaton, is not
invariant under this transformation (see eq. (\ref{Strafo})),
as naively expected, is a
typical one-loop effect in heterotic perturbation theory \cite{CDFP}.

\section{Conclusions and outlook}
\setcounter{equation}{0}
In this paper we have constructed the first explicit example of an
effective gauged supergravity Lagrangian that incorporates extra
light modes descending from branes wrapped on a vanishing cycle.
We did this for a particular situation, $SU(2)$ gauge symmetry
enhancement without charged matter in the compactification of
M-theory on a Calabi-Yau threefold. From the duality to heterotic string
theory on $K3\times S^1$, it follows that our results also apply to
stringy perturbative mechanisms of gauge symmetry enhancement.

After solving a complicated algebraic embedding problem, we
encountered a structure that is very rigid and seems to be
completely determined by gauge symmetry. The first step of
reversing the effects of integrating out charged multiplets
results in a symmetrization of the prepotential ${\cal V}$ with
respect to the transformation naturally associated with the
singularity. In our examples these were either elementary
transformations or  proper transformations in the case of the
flop. In both cases one gets the sum over orbits of the
transformation, \be \hat{\cal V}|_{{\cal M }} = \ft12 ( {\cal V} +
\tilde{\cal V} ). \ee In the second step, which we only carried out
for the specific case of $SU(2)$ gauge symmetry enhancement
without additional matter, the re-installation of the extra
mutiplets is achieved by replacing one variable, $\hat{h}^3
\propto h^1 \propto (T - U)$, by the appropriate invariant:
$(\hat{h}^3)^2 \rightarrow (\hat{h}^1)^2 +(\hat{h}^2)^2 +
(\hat{h}^3)^2 $. We expect that a similar systematics is also
present in more general cases.

The work presented here can be extended in various directions. One
main direction is the systematic construction of supergravity
Lagrangians from compactifications of M-theory on singular
manifolds. Here M-theory is understood to include perturbative
string compactifications, generalized string compactifications
with fluxes and branes, as well as F-theory. We expect that the
known relations between ungauged supergravity and smooth manifolds
extend in a systematic way to gauged supergravity and singular
manifolds. As we already mentioned, this might also be interesting
from the mathematical point of view.

Some physical applications of such constructions were already mentioned
in the introduction, but there are various other interesting
aspects that might be worth studying.
Compactifications on singular manifolds might for example
provide an alternative
tool for embedding certain  supersymmetric brane world scenarios
(such as \cite{RS,susyRS}) into string or M-theory.
The bulk theories of many of these models typically
involve certain types of 5D, $\mathcal{N}=2$ gauged supergravity theories,
and it is not always known how these can be embedded into M-theory.
One well-understood mechanism for generating gauged supergravity theories
from compactifications is to turn on background fluxes (see   e.g.
\cite{BG,LOSW}
for the particular case of  M-theory compactifications on
Calabi-Yau threefolds. For
 an exhaustive list of references on more general
Calabi-Yau compactifications with background fluxes see,   e.g.,
\cite{Haack}).
While such
background fluxes lead to rather generic gaugings of the R-symmetry group,
the gaugings of scalar manifold isometries one obtains from this mechanism
are  rather non-generic. In this sense, compactifications on
singular manifolds might be viewed as  a complementary tool for generating
 gauged supergravities from compactifications. Natural further steps
include the addition of hypermultiplets or compactifications on
singular background manifolds in combination with fluxes.

It will also be very interesting to investigate four-dimensional
$\mathcal{N}=2$ supergravity along the same lines. The simplest
case to discuss is perturbative gauge symmetry enhancement in the
heterotic string on $K3 \times T^2$. A closely related and
even more interesting case is its non-perturbative form,
described by the conifold singularity in the dual type II compactification.
Another possible  extension is to consider higher rank non-Abelian gauge
groups, which are unbroken on loci of higher codimension in the
moduli space.
A further, but more difficult
step would of course  be the extension to four-dimensional
$\mathcal{N}=1$ compactifications.

In Section \ref{ours} we mentioned an example of an  apparent space-time
singularity in a  solution of five-dimensional supergravity. This
illustrates the second main direction of application and extension
of the results of this paper: the study of space-time backgrounds
where topological phase transistions or gauge symmetry enhancement
are realized dynamically. By this we mean that in a background
with space-time dependent moduli the moduli can evolve in such a
way that they reach the special locus corresponding to  gauge
symmetry enhancement or  topological transitions. This is closely
related to the absence of space-time singularities in stringy
backgrounds, because the new dynamics occuring at special points
in moduli space are among the mechanisms which resolve or excise
space-time singularities that are apparently present in a naive
analysis based on (super)-gravity \cite{Johnson,KalMohShm}.

Presently under investigation are
BPS solutions of five-dimensional supergravity coupled
to vector multiplets,
where the moduli vary through the K\"ahler cone.
This was already discussed in subsection \ref{ours}, and the
results will be presented in a forthcoming paper \cite{TM1}.
As an extension of this one would like to investigate
the same solutions in the context of the full $SU(2)$ gauged
supergravity theory constructed in this paper. While solutions
in the ungauged effective theory are not smooth, but only
$C^2$ at the points where
the $SU(2)$ enhancement happens, one might expect that this
is an artifact of integrating out massless supermultiplets and
that the discontinuities are smoothed out when lifting the solutions
to the full gauged supergravity. Whether this really happens is not
clear to us at this point, but we expect that the comparison of the
same backgound in terms of two different effective Lagrangians
will  be very instructive.

In \cite{GSS}, the effects
of certain topological transitions within the setup of heterotic M-theory
\cite{HW} were investigated, and it would be interesting to extend
this analysis to more general types of transitions.
A further extension  along these lines could be a systematic study
of time-dependent
solutions, in particular those relevant for cosmology and the
analysis of big bang and big crunch singularities
 (for a recent discussion see e.g.    \cite{KOSST}).

\vspace{2cm}

\acknowledgments
We would like to thank J. Louis for many useful discussions.
The interest of one of the authors [T.M.] in the work presented here
resulted from inspiring discussions with S. Kachru, R. Kallosh,
M. Shmakova and E. Silverstein on the results of \cite{KalMohShm}.
M.Z.  thanks J. Ellis and M. G\"{u}naydin for discussions on
related material from  \cite{EGZ}. We would further
 like to thank B.~Szendr\H{o}i
for bringing ref. \cite{BS} to our attention, which discusses elementary
transformations from the mathematical point of view. We also thank him for 
pointing out a terminological error in the first version of the present paper.
Similarly, we thank W.~M\"{u}ck for contributing an elegant shortcut 
that considerably simplifies the first part of the calculation given
in Appendix A.

\vspace{2.5cm}

\noindent
{\bf\Large Appendix}

\begin{appendix}

\renewcommand{\theequation}{A.\arabic{equation}}
\section{The embedding}
\setcounter{equation}{0}
This appendix contains some details on the reconstruction of the polynomial
$\hat{\mathcal{V}}$.

Consider the polynomial
\begin{equation}\label{3Na}
\mathcal{V}(S,T,U)=STU+\frac{1}{3}U^{3} + a(T-U)^{3}
\end{equation}
where $a$ is an as yet arbitrary constant (The original polynomials
 considered in Section 2.3 obviously   corespond to $a=0$,
whereas the version  that takes into account the one-loop threshold effects
(Section 3.3) would correspond to $a=1/6$. We leave this constant $a$
open at this point
in order to see whether there are preferred values from the embedding
point of view.)

In the following, we will use slightly different coordinates
\begin{eqnarray}
X&=&U\\
Y&=&(T-U)\\
Z&=&S
\end{eqnarray}
in terms of which the polynomial (\ref{3Na}) reads
\begin{equation}\label{3Nb}
\mathcal{V}(X,Y,Z)=ZX^2+XYZ+\frac{1}{3}X^{3}+aY^{3}.
\end{equation}
This polynomial describes a two-dimensional scalar manifold
$\M$. The set of points given by $Y=0$ forms a codimension
one hypersurface, i.e., a line, in $\M$.

Our goal now is to embed the two-dimensional scalar manifold
$\M$ based on (\ref{3Nb}) into the four-dimensional scalar
manifold $\hat{\M}$ based on
   the    polynomial $\hat{\mathcal{V}}$ (see eq.    (\ref{5N}))
 such that the line
$Y=0$ in   $\M$ coincides with (at least part of) the line
of $SU(2)$ invariant points in $\hat{\M}$.

We assume that this embedding can be achieved by simply embedding
the corresponding polynomials  into each other.

Requiring that $Y=0$ implies ${\hat{h}}^A=0$ then leads to the most
general ansatz
(which has to be linear to keep the polynomial cubic)
\begin{eqnarray}
\hh^0&=&AX+BY+CZ\\
\hh^1&=&DY\\
\hh^2&=&EY\\
\hh^3&=&FY\\
\hh^4&=&GX+HY+IZ
\end{eqnarray}
with $R^2:= D^2+E^2+F^2\neq 0$.

Inserting this into (\ref{5N}) and comparing with the coefficients
of (\ref{3Nb}) leads to a coupled system of ten cubic equations
for the coefficients $A,\ldots,I$:
\begin{eqnarray}
A^{3}-\frac{3}{2}AG^2+\lambda G^3&=&\frac{1}{3}\label{1}\\
B^{3}-\frac{3}{2}BH^2+\lambda H^3 - \frac{3}{2}R^2[B+\kappa
H]&=&a\label{2}\\
 C^{3}-\frac{3}{2}CI^2+\lambda I^3&=&0\label{3}\\
B \left[ 3A^2-\frac{3}{2}G^2\right] +H\left[-3AG+3\lambda G^2
\right]&=&0\label{4}\\
 C \left[ 3A^2-\frac{3}{2}G^2\right]
+I\left[-3AG+3\lambda G^2 \right]&=&1\label{5}\\
A\left[3B^2-\frac{3}{2}R^2-\frac{3}{2}H^2\right]+G\left[-3BH+3\lambda
H^2-\frac{3}{2}\kappa R^2\right]&=&0\label{6}\\
C\left[3B^2-\frac{3}{2}R^2-\frac{3}{2}H^2\right]+I\left[-3BH+3\lambda
H^2-\frac{3}{2}\kappa R^2\right]&=&0\label{7}\\
A\left[3C^2-\frac{3}{2}I^2\right]+G\left[-3CI+3\lambda
I^2\right]&=&0\label{8}\\
B\left[3C^2-\frac{3}{2}I^2\right]+H\left[-3CI+3\lambda
I^2\right]&=&0\label{9}\\
 6ABC-3[AHI+GIB+CGH]+6\lambda GHI&=&1\label{10}
\end{eqnarray}

These equations also involve the unknown coefficients
$(\lambda,\kappa)$ as well as $a$. Being optimistic, one might
therefore hope that a consistent solution to the above equations
only exists for a small number of values for $(\lambda,\kappa)$
and $a$. We will now see to what extent this is the case. Our strategy is to
   first  show    that
a consistent solution of (\ref{1})-(\ref{10}) can only exist if
$\lambda^2=\frac{1}{2}$. As mentioned in Section 3.2, a  redefinition
${\hat{h}}^4\rightarrow -{\hat{h}}^4$ can always be used to render $\lambda$
positive, i.e., we can always choose the positive root
$\lambda=+\frac{1}{\sqrt{2}}$. Let us sketch how this result is
obtained.\footnote{This more elegant derivation of $\lambda^2=\frac{1}{2}$
replaces a lengthier
argument given in the first version of this paper and was contributed by 
W.~M\"{u}ck.}

First, rewrite eqs. (\ref{3}) and (\ref{8}) as
\begin{eqnarray}
C[C^2-\frac{1}{2}I^2]+I[-CI+\lambda I^2]&=&0\\
A[C^2-\frac{1}{2}I^2]+G[-CI+\lambda I^2]&=&0.
\end{eqnarray}
Viewing this as a linear system for $[C^2-\frac{1}{2}I^2]$ and
$[-CI+\lambda I^2]$, one sees that either $[C^2-\frac{1}{2}I^2]=
[-CI+\lambda I^2]=0$, or $CG-AI=0$. 

Let us first assume $CG-AI=0$. Multiplying (\ref{1}) by $I^3$,
one then derives $I=0$ by taking into account eq. (\ref{3}).
In a similar way, one derives $C=0$ by multiplying (\ref{1})
by $C^3$. However, $C=I=0$ is a contradiction to (\ref{5}).
Hence,  we have to assume the second possibility $[C^2-\frac{1}{2}I^2]=
[-CI+\lambda I^2]=0$. But then, because $C=I=0$ is ruled out by
(\ref{5}), we immediately find the desired result $\lambda^2=\frac{1}{2}$.

To sum up, we have shown that if a consistent embedding exists, it
can at most work with
\begin{equation}\label{lambda2}
\lambda^2=\frac{1}{2}.
\end{equation}
As mentioned above, there is no loss of generality in choosing
the positive root
\begin{equation}\label{lambda}
\lambda=+\frac{1}{\sqrt{2}}
\end{equation}
as we will do from now on.

Note that we have never used (\ref{2}) to arrive at this result.
(\ref{2}) is the only equation that involves the parameter $a$,
which implies that our conclusion regarding $\lambda$ is valid for
any $a$.

As our second step, we will now use (\ref{lambda2}) to simplify
our original polynomial (\ref{5N}). To be explicit,
(\ref{lambda2}) implies that, in terms of the new coordinates
\begin{eqnarray}
w&:=&\hh^0+2\lambda \hh^{4}\\
v&:=&\hh^0-\lambda \hh^{4},
\end{eqnarray}
the polynomial (\ref{5N}) simplifies to
\begin{equation}\label{5Nb}
\hat{\mathcal{V}}(w,v,\hh^A)=wv^2 -
\frac{1}{2}\left[w\left(1+\frac{\kappa}{\lambda}\right)+
v\left(2-\frac{\kappa}{\lambda}\right)\right] \delta_{AB}\hh^A \hh^B.
\end{equation}

We will now forget about the original form (\ref{5N}), as well as
the corresponding embedding equations (\ref{1})-(\ref{10}).
Instead, we will now exclusively use the new form (\ref{5Nb}),
because the corresponding embedding equations are much easier to
solve.

That is, we  now parameterize the  embedding  as
\begin{eqnarray}
w&=&AX+BY+CZ\\
\hh^1&=&DY\\
\hh^2&=&EY\\
\hh^3&=&FY\\
v&=&GX+HY+IZ
\end{eqnarray}
with $(D^2+E^2+F^2)\neq 0$. Note that the coefficients here have nothing
to do with the old coefficients $A,\ldots,I$ we used earlier.

Inserting now this ansatz into (\ref{5Nb}) and comparing with
(\ref{3Nb}) yields the new embedding conditions
\begin{eqnarray}
AG^2&=&\frac{1}{3}\label{q1}\\
BH^2-\frac{1}{2}R^2\left[B\left(1+\frac{\kappa}{\lambda}\right)+H\left(2-
\frac{\kappa}{\lambda}\right)\right]&=&a\label{q2}\\
CI^2&=&0\label{q3}\\
AH^2+2BGH-\frac{1}{2}R^2\left[A\left(1+\frac{\kappa}{\lambda}\right)+G\left(2-
\frac{\kappa}{\lambda}\right)\right]&=&0\label{q4}\\
AI^2+2CGI&=&0\label{q5}\\
BG^2+2AGH&=&0\label{q6}\\
CG^2+2AGI&=&1\label{q7}\\
BI^2+2CHI&=&0\label{q8}\\
CH^2+2BHI-\frac{1}{2}R^2\left[C\left(1+\frac{\kappa}{\lambda}\right)+I\left(2-
\frac{\kappa}{\lambda}\right)\right]&=&0\label{q9}\\
2[AHI+BGI+CGH]&=&1\label{q10},
\end{eqnarray}
where again $R^2\equiv D^2 +E^2+ F^2$.  We will now solve these
equations. Our strategy is to  first solve $A,B,C,H,I$ in terms of
$G$ and then relate $G$ to the other parameters.
We start from eq.
(\ref{q1}), which implies $A,G\neq 0$ and
\begin{equation}
A=\frac{1}{3G^2}.
\end{equation}
$C$ can be either zero or not. If $C=0$, (\ref{q5}) implies
(because of $A\neq 0$) that $I=0$. However, (\ref{q10}) forbids
that $C$ and $I$ vanish simultaneously, so that we have to assume
$C\neq 0$. But then, (\ref{q3}) tells us that
\begin{equation}
I=0.
\end{equation}
This then automatically solves (\ref{q5}) and (\ref{q8}). From
(\ref{q7}) we obtain
\begin{equation}
C=\frac{1}{G^2}.
\end{equation}
(\ref{q10}) then implies
\begin{equation}
H=\frac{G}{2}.
\end{equation}
From eq. (\ref{q6}), we then obtain
\begin{equation}
B=-\frac{1}{3G^2}
\end{equation}
Eq. (\ref{q9}) implies
\begin{equation}\label{R1}
R^2\left(1+\frac{\kappa}{\lambda}\right)=2H^2,
\end{equation}
which gives us a constraint on $\kappa$:
\begin{equation}\label{k1}
\kappa>-\lambda.
\end{equation}
From eqn. (\ref{q4}), we learn that
\begin{equation}\label{R2}
R^2\left(2-\frac{\kappa}{\lambda}\right)=-\frac{2}{3G},
\end{equation}
which gives us another constraint on $\kappa$:
\begin{equation}\label{k2}
\kappa\neq 2\lambda
\end{equation}
as well as the consistency conditions (because of $R^2>0$)
\begin{eqnarray}
G>0&\Leftrightarrow& \kappa>2\lambda\label{k3}\\
G<0&\Leftrightarrow& \kappa<2\lambda\label{k4}.
\end{eqnarray}
Adding (\ref{R1}) and (\ref{R2}) gives $R^2$ in terms of $G$:
\begin{equation}\label{R3}
R^2=\frac{G^2}{6}-\frac{2}{9G}
\end{equation}
whereas dividing (\ref{R1}) by (\ref{R2}) (which is always
possible because of $R^2\neq 0$ and (\ref{k2})) yields $G$ in terms
of $\kappa$:
\begin{equation}\label{Gk}
G^3=-\frac{4(\lambda+\kappa)}{3(2\lambda-\kappa)}.
\end{equation}
One now convinces oneself that any value for $\kappa$ in the
allowed region (cf. eqs. (\ref{k1}), (\ref{k2}))
\begin{equation}
\kappa\in (-\lambda,2\lambda)\cup (2\lambda,+\infty)
\end{equation}
gives a value for $G$ via (\ref{Gk}) that automatically also
satisfies the consistency conditions (\ref{k3})-(\ref{k4}) and
gives rise to a positive value for $R^2$ via (\ref{R3}). The other
coefficients $A,B,C,H,I$ can then be unambiguously calculated from
$G$ using the above equations.

At this point,  we have therefore   shown that any $\kappa\in
(-\lambda,2\lambda)\cup (2\lambda,+\infty)$ leads to a consistent
solution to eq. (\ref{q1}) and eqs. (\ref{q3})-(\ref{q10}). The
only equation we have not yet used/solved, is equation
(\ref{q2}), which is the only equation involving the parameter
$a$. Thus, all we have derived so far applies to all possible $a$.
Inserting our results into eq. (\ref{q2}), however, fixes $a$ to
be
\begin{equation}
a=\frac{1}{6}
\end{equation}
without that any new constraints are imposed on the other
coefficients. Remarkably, $a=\frac{1}{6}$ exactly corresponds to
the value that  is  favored if one takes the one-loop threshold effects into
account (see Section 3.3).

Putting everything together, the embedding of (\ref{3Nb}) into
(\ref{5N}) fixes $\lambda$ and $a$, and restricts $\kappa$ to take on
values in two intervals:
\begin{eqnarray}
a&=&\frac{1}{6}\\
\lambda&=&\frac{1}{\sqrt{2}}\\
\kappa&\in&(-\frac{1}{\sqrt{2}},\sqrt{2})\cup(\sqrt{2},+\infty).
\end{eqnarray}

\renewcommand{\theequation}{B.\arabic{equation}}
\section{All the allowed $\kappa$ lead to equivalent
$\hat{\M}$}
\setcounter{equation}{0}

Naively, one might now conclude that the extended scalar manifold
 $\hat{\M}$ is not unique, and  that, instead,  there is a
one-parameter family of such manifolds parameterized by the
allowed values for $\kappa$. We will now show that this is not
true, because all the allowed $\kappa$ lead to equivalent
manifolds.

To be precise, we will show that  any  $\kappa_{1}\in
(-\lambda,2\lambda)\cup (2\lambda,+\infty)$ can be transformed
into any other  $\kappa_{2}\in (-\lambda,2\lambda)\cup
(2\lambda,+\infty)$ by means of a simple rescaling of the
variables $(v,w,\hh^{A})$ of the polynomial (\ref{5Nb})

To see this, choose two arbitrary allowed $\kappa$ values
$\kappa_{1},\kappa_{2}\in (-\lambda,2\lambda)\cup
(2\lambda,+\infty)$.

Then define
\begin{eqnarray}
\rho&:=&\left(\frac{\lambda+
\kappa_{1}}{\lambda+\kappa_{2}}\right)^{\frac{1}{6}}
\left(\frac{2\lambda-\kappa_1}{2\lambda-\kappa_{2}}\right)^{\frac{1}{3}}\\
\sigma&:=&\left(\frac{2\lambda-\kappa_2}{2\lambda-\kappa_{1}}\right)\rho^2.
\end{eqnarray}
Note that $(\lambda+\kappa_i)>0$ and $\kappa_{i}\neq 2\lambda$, so
that $\rho$ and $\sigma$ are well-defined, non-vanishing real
numbers. Now consider the polynomial (\ref{5Nb}) with
$\kappa=\kappa_{1}$,
\begin{equation}
\hat{\mathcal{V}}(w,v,\hh^A)=wv^2 -
\frac{1}{2}\left[w\left(1+\frac{\kappa_1}{\lambda}\right)+
v\left(2-\frac{\kappa_1}{\lambda}\right)\right] \delta_{AB}\hh^A
\hh^B.
\end{equation}

After the  coordinate  rescalings
\begin{eqnarray}
v& \longrightarrow& \sigma v\\
w& \longrightarrow& \sigma^{-2} w\\
\hh^{A}&   \longrightarrow& \rho^{-1} \hh^{A}
\end{eqnarray}
this becomes
\begin{equation}
\hat{\mathcal{V}}(w,v,\hh^A)=wv^2 -
\frac{1}{2}\left[w\left(1+\frac{\kappa_2}{\lambda}\right)+
v\left(2-\frac{\kappa_2}{\lambda}\right)\right] \delta_{AB}\hh^A
\hh^B,
\end{equation}
which is nothing but (\ref{5Nb}) with $\kappa=\kappa_2$. Thus all
our polynomials (\ref{5Nb}) with $\kappa\in
(-\lambda,2\lambda)\cup (2\lambda,+\infty)$ are equivalent and
describe the same scalar manifold $\hat{\M}$, which is thus
unique.

\end{appendix}

\end{document}